\documentclass[12pt,preprint]{aastex}
\usepackage{emulateapj5,apjfonts}
\usepackage{graphics}
\usepackage{amssymb}
\usepackage{onecolfloat}
\usepackage{bm}
\lefthead{Luki\'c, Heitmann, Habib, Bashinsky, Ricker}
\righthead{The Halo Mass Function}

\newcommand{\mf}{F}
\newcommand{\tvph}{\vphantom{$\sqrt{\frac{2a}{\pi}}$}}

\begin{document}


\def\head{
  \vbox to 0pt{\vss
                    \hbox to 0pt{\hskip 440pt\rm LA-UR-06-0847\hss}
                   \vskip 25pt}

\title{The Halo Mass Function: High--Redshift Evolution and Universality}
\author{Zarija~Luki\'c\altaffilmark{1}, Katrin~Heitmann\altaffilmark{2},
        Salman~Habib\altaffilmark{3}, Sergei
        Bashinsky\altaffilmark{3}, and         
        Paul~M.~Ricker\altaffilmark{1,4}}

\affil{$^1$ Dept.\ of Astronomy, University of Illinois, Urbana, IL 61801}
\affil{$^2$ ISR-1, ISR Division,  Los
Alamos National Laboratory, Los Alamos, NM 87545}
\affil{$^3$ T-8, Theoretical Division, Los
Alamos National Laboratory, Los Alamos, NM 87545}
\affil{$^4$ National Center for Supercomputing Applications, Urbana, IL 61801}

\date{today}

\begin{abstract}
  
  We study the formation of dark matter halos in the concordance
  $\Lambda$CDM model over a wide range of redshifts, from $z=20$ to
  the present. Our primary focus is the halo mass function, a key
  probe of cosmology.  By performing a large suite of nested-box
  $N$-body simulations with careful convergence and error controls (60
  simulations with box sizes from 4 to 256$\,h^{-1}$Mpc),
  we determine the mass function and its evolution with excellent
  statistical and systematic errors, reaching a few percent over most
  of the considered redshift and mass range. Across the studied
  redshifts, the halo mass is probed over 6 orders of magnitude
  ($10^7$ -- $10^{13.5}\,h^{-1}M_\odot$).  Historically, there has
  been considerable variation in the high redshift mass function as
  obtained by different groups. We have made a concerted effort to
  identify and correct possible systematic errors in computing the
  mass function at high--redshift and to explain the discrepancies
  between some of the previous results. We discuss convergence
  criteria for the required force resolution, simulation box size,
  halo mass range, initial and final redshifts, and time
  stepping. Because of conservative cuts on the mass range probed by
  individual boxes, our results are relatively insensitive to
  simulation volume, the remaining sensitivity being consistent with
  extended Press-Schechter theory. Previously obtained mass function
  fits near $z=0$, when scaled by linear theory, are in good agreement
  with our results at all redshifts, although a mild redshift
  dependence consistent with that found by Reed et al. may exist at
  low redshifts. Overall, our results are consistent with a
  ``universal'' form for the mass function at high redshifts.

\end{abstract}

\keywords{methods: $N$-body simulations ---
          cosmology: halo mass function}}

\twocolumn[\head]
\section{Introduction}

A broad suite of astrophysical and cosmological observations provides
compelling evidence for the existence of dark matter. Although its
ultimate nature is unknown, the large-scale dynamics of dark matter is
essentially that of a self-gravitating collisionless fluid. In an
expanding universe, gravitational instability leads to the
formation and growth of structure in the dark matter distribution. The
existence of localized, highly overdense dark matter clumps, or halos, 
is a key prediction of cosmological nonlinear gravitational
collapse. The distribution of dark matter halo masses is termed the
halo mass function and constitutes one of the most important probes of
cosmology.  At low redshifts, $z\leq 2$, the mass function at the
high-mass end (cluster scales) is very sensitive to variations in
cosmological parameters, such as the matter content of the Universe
$\Omega_{\rm m}$, the dark energy content along with its
equation-of-state parameter,~$w$ \citep{Holder01}, and the
normalization of the primordial fluctuation power spectrum,
$\sigma_8$.  At higher redshifts, the halo mass function is important
in probing quasar abundance and formation sites \cite[]{Haiman01}, as
well as the reionization history of the Universe~\cite[]{Furl06}.

Many recently suggested reionization scenarios are based on the
assumption that the mass function is given reliably by modified
Press-Schechter type fits (Press \& Schechter 1974, 
hereafter PS; Bond et al. 1991).
However, the theoretical basis of this approach is at best heuristic
and careful numerical studies are required in order to obtain accurate
results. Two examples serve to illustrate this statement. Reed et
al.~(2003) report a discrepancy with the Sheth-Tormen fit
(Sheth \& Tormen 1999, hereafter ST) 
of $\sim$50\% at a redshift of $z=15$ (we explain the
different fitting formulae and their origin in
\S \ref{massf}). Heitmann et al.~(2006a) show that the
Press-Schechter form can be severely incorrect at high redshifts: at
$z\ge 10$, the predicted mass function sinks below the numerical
results by an order of magnitude at the upper end of the relevant mass
scale. Consequently, incorrect, or at best imprecise, predictions for the
reionization history can result from the failure of fitting formulae.

Since halo formation is a complicated nonlinear gravitational process,
the current theoretical understanding of the mass, spatial
distribution, and inner profiles of halos remains at a relatively
crude level. Numerical simulations are therefore crucial as drivers of
theoretical progress, having been instrumental in obtaining important
results such as the Navarro-Frenk-White (NFW) profile \cite[]{NFW97}
for dark matter halos and an (approximate) universal form for the mass
function (Jenkins et al.\, 2001, hereafter Jenkins). In order to better understand the
evolution of the mass function at high redshifts, a number of
numerical studies have been carried out. High--redshift simulations,
however, suffer from their own set of systematic issues, and
simulation results can be at considerable variance with each other,
differing on occasion by as much as an order of magnitude!

Motivated by all of these reasons we have carried out a numerical
investigation of the evolution of the mass function with the aim of
attaining good control over both statistical and, more importantly,
possible systematic errors in $N$-body simulations. Our first results
have been reported in condensed form in Heitmann et al. (2006a). Here
we provide a more detailed and complete exposition of our work,
including several new results.

We first pay attention to simulation criteria for obtaining accurate
mass functions with the aim of reducing systematic effects. 
Our two most significant points are that simulations must be started early
enough to obtain accurate results 
and that the box sizes must be large enough
to suppress finite-volume artifacts. As in most recent work following
that of Jenkins, we define halo masses using a
friends-of-friends (FOF) halo finder with linking length $b=0.2$. This
choice introduces systematic issues of its own (e.g., connection to
spherical overdensity mass as a function of redshift), which we touch on 
as relevant below. As it is not quantitatively significant in the
context of this paper, we leave a detailed discussion to later work
(Z. Luki\'c et al., in preparation; see also \citealt{Reed07}).

The more detailed results in this paper enable us to study the mass
function at statistical and systematic accuracies
reaching a few percent over most of our redshift range, a substantial
improvement over most previous work. At this level we find
discrepancies with the ``universal'' fit of Jenkins at
low redshifts ($z<5$), but it must be kept in mind that the
universality of the original fit was only meant to be at the $\pm
20\%$ level. Recently, \cite{Reed07} have found violation of
universality at high redshifts (up to $z=30$). To fit the mass
function they have incorporated an additional free parameter, the
effective spectral index $n_{\rm eff}$, with the aim of understanding
and taking into account the extra redshift dependence missing from
conventional mass--function--fitting formulae. Our simulation results
are consistent with the trends found by \cite{Reed07} at low redshifts
($z\leq 5$), but at higher redshifts we do not observe a statistically
significant violation of the universal form of the mass function.

Results from some previous simulations have reported good agreement
with the Press-Schechter mass function at high redshifts.  Since the
Press-Schechter fit has been found significantly discrepant with 
low--redshift results ($z<5$), this would imply a strong disagreement
with extending the well-validated low--redshift notion of (approximate)
mass function universality to high~$z$.  Our conclusion is that the
simulations on which these findings were based violated one or more of
the criteria to be discussed below.

As simulations are perforce restricted to finite volumes, the obtained
mass function clearly cannot represent that of an infinite box. Not
only is sampling a key issue, but also the fact that simulations with
periodic boundary conditions have no fluctuations on scales larger
than the box size. To minimize and test for these effects we were
conservative in our choices of box size and the mass range probed in
each individual box. We also used nested-volume simulations to
directly test for finite-volume effects. Because we used multiple
boxes and averaged mass function results over the box ensemble,
extended Press-Schechter theory can be used to correct for residual
finite volume--effects (\citealt{Mo96}; \citealt{Barkana04}); this
approach is different from the individual box corrections applied by
\cite{Reed07}. Details are given in \S \ref{simvol}.

The paper is organized as follows.  In \S \ref{massf} we give a
brief overview of the mass function and popular fitting formulae,
discussing as well previous numerical work on the halo mass function
at high redshifts.  In \S \ref{code} we give a short description
of the $N$-body code MC$^2$ ({\bf M}esh-based {\bf C}osmology {\bf
  C}ode) and a summary of the performed simulations. In
\S \ref{icevol} we derive and discuss some simple criteria for
the starting redshift and consider systematic errors related to the
numerical evolution such as mass and force resolution and time
stepping. These considerations in turn specify the input parameters
for the simulations in order to span the desired mass and redshift
range for our investigation. In \S\ref{resint} we present
results for the mass function at different redshifts as well as the
halo growth function. Here we also discuss the importance of
post-processing corrections such as FOF particle sampling compensation
and finite-volume effects. We discuss our results and conclude in
\S\ref{conclusion}.

\section{Definitions and Previous Work}
\label{massf}

The mass function describes the number density of halos of a given
mass. In order to determine the mass function in simulations one has
to first {\em identify} the halos and then {\em define} their mass. No
precise theoretical basis exists for these operations. Nevertheless,
depending on the situation at hand, the observational and numerical
communities have adopted a few ``standard'' ways of defining halos and
their associated masses.  For a recent review of these issues with
regard to observations, see, e.g., ~\cite{Voit05}, but for a more
theoretically oriented review, see, e.g.,~\cite{White01}.

\subsection{Halo Mass}
\label{halomass}

There are basically two ways to find halos in a simulation.  One, the
overdensity method, is based on identifying overdense regions above a
certain threshold. The threshold can be set with respect to the
critical density $\rho_{\rm c}=3H^2/8\pi G$ (or the background density
$\rho_{\rm b}=\Omega_{\rm m}\rho_{\rm c}$, where $\Omega_{\rm m}$ is
the matter density of the Universe including dark matter and
baryons). The mass $M_\Delta$ of a halo identified this way is defined
as the mass enclosed in a sphere of radius $r_\Delta$ whose mean
density is $\Delta\rho_{\rm c}$. Common values for $\Delta$ range
from 100 to 500 (or even higher).  As explained in \cite{Voit05},
cluster observers prefer higher values for $\Delta$. Properties of
clusters are easier to observe in higher density regions and these
regions are more relaxed than the outer parts which are subject to the
effects of inflow and incomplete mixing. The disadvantage of defining
a halo in this manner is that sphericity of halos is implied, an
assumption which may be easily violated, e.g., in the case of halos
that formed in a recent merger event or halos at high redshifts.
At higher redshifts, the nonlinear mass scale $M_*$ decreases rapidly, 
and the ratio of the considered halo mass $M_{\rm halo}$ to $M_*$ can
become large. This translates into producing large-scale structures
roughly analogous to supercluster structures today. While these
structures are gravitationally bound, they are often not virialized,
nor spherical. Even the much smaller structures (which are considered
in this paper) are not virialized at high redshifts, and therefore, 
assumptions about sphericity are most likely violated.  Hence the
spherical overdensity method does not suggest itself as an obvious way
to identify halos at high redshift.

The other method, the FOF algorithm, is based on
finding neighbors of particles and neighbors of neighbors as defined
by a given separation distance (see, e.g., Einasto et al.~1984; Davis
et al.~1985).  The FOF algorithm leads to halos with arbitrary shapes
since no prior symmetry assumptions have been made.  The halo mass is
defined simply as the sum of particles which are members of the halo.
While this definition is easy to apply to simulations, the connection
to observations is difficult to establish directly.  (For an
investigation of connections between different definitions of halos
masses and approximate conversions between them, see White~2001).
 
It is important to keep in mind that the definition of a halo is
essentially the adoption of some sort of convention for the halo
boundary. In reality, a sharp distinction between the particles in a
halo and particles in the simulation ``field'' does not
exist. Jenkins showed that the choice of a FOF finder with a linking length $b=0.2$ 
to define halo masses provides the best fit for a universal form of
the mass function. This choice has since been adopted by many
numerical practitioners as a standard convention.  A useful discussion
of the various halo definitions can be found in \cite{White02}.

In this paper we use the FOF algorithm to identify halos and their
masses. It was recently pointed out by Warren et al.~(2006, hereafter Warren) 
that FOF masses
suffer from a systematic problem when halos are sampled by relatively
small numbers of particles. Although halos can be robustly identified
with as few as 20 particles, if a given halo has too few particles,
its FOF mass turns out to be systematically too high. We describe how
we compensate for this effect in \S\ref{masscorr}.  In the
current paper, all results for the mass function are displayed at a
fixed FOF linking length of $b=0.2$, using the Warren 
correction.

\subsection{Defining the Mass Function}
\label{massdef}

The exact definition of the mass function, e.g., integrated versus
differential form or count versus number density, varies widely in
the literature.  To characterize different fits, Jenkins 
introduced the scaled differential mass function $f(\sigma, z)$ as
a fraction of the total mass per $\ln\sigma^{-1}$ that belongs to
halos:
\begin{equation}
\label{fsigma}
f(\sigma, z) \equiv \frac{d\rho/\rho_b}{d\ln\sigma^{-1}}
                   =\frac{M}{\rho_{\rm b}(z)} \frac{dn(M,z)}
                    {d\ln[\sigma^{-1}(M,z)]}.
\end{equation}
Here $n(M,z)$ is the number density of halos with mass~$M$,
$\rho_{\rm b}(z)$ is the background density at redshift $z$, and
$\sigma(M,z)$ is the variance of the linear density field.  As pointed
out by Jenkins, this definition of the mass function
has the advantage that to a good accuracy
it does not explicitly depend on redshift,
power spectrum, or cosmology; all of these are encapsulated in
$\sigma(M,z)$. For the most part, we will display the mass
function 
\begin{equation}
\label{FM}
\mf(M,z) \equiv {dn \over d\log M}
\end{equation} 
as a function of $\log M$ itself. [In \S\ref{resint} we 
include results for $f(\sigma,z)$.] 

To compute $\sigma(M,z)$, the
power spectrum $P(k)$ is smoothed with a spherical top-hat filter
function of radius $R$, which on average encloses a mass $M$ ($R =
[3M/4\pi \rho_{\rm b}(z)]^{1/3}$):
\begin{equation}
\sigma^2(M,z) = 
\frac{d^2(z)}{2\pi^2}\int^{\infty}_{0}k^2P(k)W^2(k,M)dk,
\label{sig}
\end{equation}
where $W(k,M)$ is the top-hat filter:
\begin{eqnarray}
W(r) & = & \left\{ \begin{array}{rl}
                            \frac{3}{4 \pi R^3}, & r<R \\
                             0,                          & r>R
                            \end{array} \right.\\
W(k) & = & \frac{3}{(kR)^3} \left[ \sin (kR) - kR \cos (kR) \right].
\end{eqnarray}
The redshift dependence enters only through the growth factor $d(z)$,
normalized so that $d(0)=1$:
\begin{equation}
\sigma(M,z) = \sigma(M,0) d(z).
\end{equation}
In the approximation of negligible difference in the CDM and baryon
peculiar velocities, the growth function in a $\Lambda$CDM universe is
given by (Peebles 1980)
\begin{equation}
d(a) = \frac{D^{+}(a)}{D^{+}(a=1)},
\end{equation}
where we consider~$d$ as a function of the cosmological scale factor
$a=1/(1+z)$, and
\begin{equation}
D^+(a)=\frac{5\Omega_{\rm m}}{2}\,\frac{H(a)}{H_0}
       \int_0^a\frac{da'}{[a'H(a')/H_0]^3}
\label{D_no_baryons}
\end{equation}
with $H(a)/H_0= \left[\Omega_{\rm m}/a^3+(1-\Omega_{\rm
    m})\right]^{1/2}$.  In particular, for $z\gg1$, when matter dominates
the cosmological constant, $D^+(a)\simeq a$.

Even in linear theory, equation~(\ref{D_no_baryons}) is only an
approximation because baryons began their gravitational collapse with
velocities different from those of CDM particles. Until recombination
at $z\sim 1100$, well into the matter era with non-negligible growth
of CDM inhomogeneities, the baryons were held against collapse by the
pressure of the CMB photons (see, e.g.~\citet{husug96}).  While
thereafter the relative baryon-CDM velocity decayed as $1/a$, the
residual velocity difference was sufficient to affect the growth
function~$d(z)$ at $z=50$ by more than $1\%$ and at $z=10$ by about
$0.2\%$ (\citealt{Yoshida03bar}; \citealt{Naoz06}).

\subsection{Fitting Functions}
\label{fits}
Over the last three decades several different fitting forms for the
mass function have been suggested. The mass function is not only a
sensitive measure of cosmological parameters by itself but also a
key ingredient in analytic and semianalytic modeling of the dark
matter distribution, as well as of several aspects of the formation,
evolution, and distribution of galaxies.  Therefore, if a reliable and
accurate fit for the mass function applicable to a wide range of
cosmologies and redshifts were to exist, it would be of obvious
utility. In this section we briefly review the common fitting
functions and compare them at different redshifts.

The first analytic model for the mass function was developed by 
PS. Their theory accounts for a
spherical overdense region in an otherwise smooth background density
field, which then evolves as a Friedmann universe with a positive
curvature.  Initially, the overdensity expands, but at a slower rate
than the background universe (thus enhancing the density contrast),
until it reaches the `turnaround' density, after which collapse
begins. Although from a purely gravitational standpoint this collapse
ends with a singularity, it is assumed that in reality -- due to the
spherical symmetry not being exact -- the overdense region will
virialize.  For an Einstein-de Sitter universe, the density of such an
overdense region at the virialization redshift is $z\approx 180
\rho_{\rm c}(z)$.  At this point, the density contrast from the linear
theory of perturbation growth [$\delta(\vec{x},z) = d(z)
\delta(\vec{x},0)$] would be $\delta_{\rm c}(z) \approx 1.686$ in an
Einstein-de Sitter cosmology.  For $\Omega_{\rm m} < 1$, the value of
the threshold parameter $\delta_{\rm c}$ can vary (see Lacey \&
Cole~1993), but the dependence on cosmology has little quantitative
significance (see, e.g., Jenkins).  Thus, throughout this
paper we adopt $\delta_{\rm c} = 1.686$.

\begin{table*}
\begin{center}
\caption{\label{tabone} Mass Function Fits for $f(\sigma)$}
\begin{tabular}{llcc}
\tableline\tableline
\qquad Reference  & \qquad Fitting Function $f(\sigma)$& Mass Range &  Redshift range \\  
\hline
ST, Sheth \& Tormen (2001) & $0.3222 \sqrt{\frac{2a}{\pi}}
\frac{\delta_{\rm c}}{\sigma} \exp  \left[ - \frac{a \delta^2_{\rm
      c}}{2\sigma^2} \right] \left[ 1 + \left( \frac{\sigma^2}{a
      \delta^2_{\rm c}} \right) ^ p \right]$ & unspecified &
unspecified\\ 

\tvph Jenkins &$ 0.315 \exp
\left[ -| \ln \sigma^{-1} + 0.61 |^{3.8} \right]$ & $-1.2\le\ln\sigma^{-1}\le1.05$ & $z=0-5$\\

\tvph Reed et al.~(2003) &$ f_{\rm ST}(\sigma)\,
\exp\left\{-0.7/\left[\sigma(\cosh(2\sigma))^5\right]\right\}$& $-1.7\le
\ln\sigma^{-1}\le0.9$ & $z=0-15$ \\

\tvph Warren & $0.7234 \left( \sigma^{-1.625} + 0.2538 \right)
\exp \left[ -\frac{1.1982}{\sigma^2}\right]$ &
$(10^{10}-10^{15}) h^{-1}M_{\odot} $ & $z=0$\\

Reed et al.~(2007) & $A \sqrt{\frac{2a}{\pi}} \left[1 +
\left(\frac{\sigma^2}{a\delta_c^2}\right)^p+0.6G_1(\sigma)+0.4G_2(\sigma)\right]$ & $-0.5\le\ln\sigma^{-1}\le1.2$ &
$z=0-30$ \\
& $ \times\frac{\delta_c}{\sigma}
\exp\left[-\frac{ca\delta_c^2}{2\sigma^2}
-\frac{0.03}{(n_{\rm eff}+3)^2}
\left(\frac{\delta_c}{\sigma}\right)^{0.6}\right]$ & \\
\tableline\tableline

\vspace{-1.5cm}

%

\tablecomments{Shown are examples of commonly used fitting functions. 
ST used $a=0.707$ and $p=0.3$, while \cite{Sheth02} 
suggest that $a=0.75$ leads to a better fit. 
The Warren fit represents by far the largest uniform set of
simulations based on multiple boxes with the same cosmology run with
the same code. We use it as a reference standard throughout
this paper. \cite{Reed03} suggest an empirical adjustement of the ST
fit, which is slightly modified in \cite{Reed07}. For the latter, 
$G_1(\sigma)$ and
$G_2(\sigma)$ are given by eqs.~(\ref{g1_def}) and~(\ref{g2_def}), 
respectively, 
$c=1.08$, $ca=0.764$, and $A=0.3222$.}
\end{tabular}
\end{center}
\end{table*}

Following the above reasoning and with the assumption that the initial
density perturbations are described by a homogeneous and isotropic
Gaussian random field, the PS mass function is specified by
\begin{equation}
f_{\rm PS}(\sigma) = \sqrt{\frac{2}{\pi}} \frac{\delta_{\rm c}}{\sigma} 
\exp \left( - \frac{\delta^2_{\rm c}}{2\sigma^2} \right).
\end{equation}
The PS approach assumes that all mass is inside halos, as enforced by
the constraint
\begin{equation}
\int^{+\infty}_{-\infty} f_{\rm PS}(\sigma)\, d\ln \sigma^{-1} = 1.
\end{equation}
While as a first rough approximation the PS mass function agrees with
simulations at $z=0$ reasonably well, it overpredicts the number of
low--mass halos and underpredicts the number of massive halos at the
current epoch. Furthermore, it is significantly in error at high
redshifts (see, e.g., Springel et al.~2005; Heitmann et al.~2006a; 
\S\ref{timeevo}).

After PS, several suggestions were made in order to improve the mass
function fit. These suggestions were based on more refined dynamical
modeling, direct fitting to simulations, or a combination of the
two.

Using empirical arguments ST 
proposed an improved mass function fit of the form:
\begin{equation}
f_{\rm ST}(\sigma) = 0.3222 \sqrt{\frac{2a}{\pi}} \frac{\delta_{\rm
    c}}{\sigma} \exp  \left( - \frac{a \delta^2_{\rm c}}{2\sigma^2} \right)
\left[ 1 + \left( \frac{\sigma^2}{a \delta^2_{\rm c}} \right) ^ p \right],
\end{equation}
with $a=0.707$ and $p=0.3$. (Sheth \& Tormen 2002 suggest $a=0.75$ as
an improved value.) Sheth et al. (2001) rederived this fit
theoretically by extending the PS approach to an elliptical collapse
model.  In this model, the collapse of a region depends not only on
its initial overdensity but also on the surrounding shear field.  The
dependence is chosen such that it recovers the Zel'dovich
approximation \cite[]{Zeldovich70} in the linear regime.  A halo is
considered virialized when the third axis collapses (see also Lee \&
Shandarin (1998) for an earlier, different approach to the same idea).

\begin{figure}[t]
  \plotone{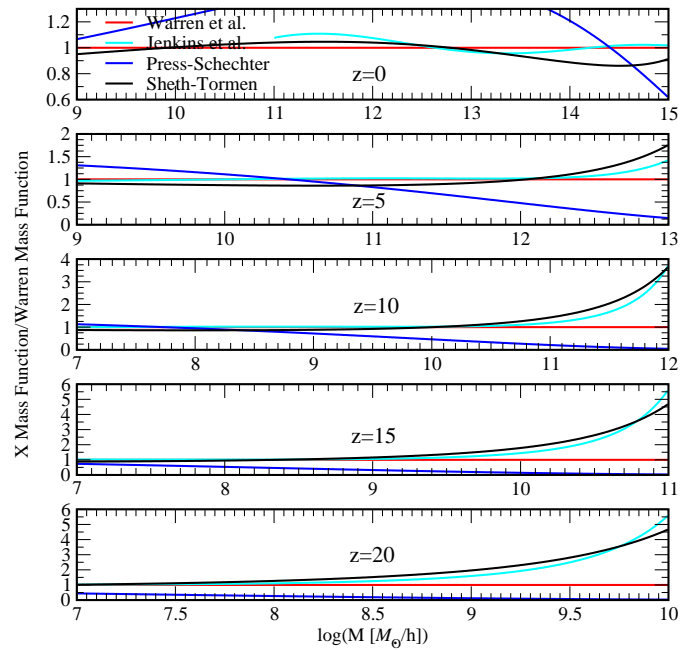}
\caption{Ratio of the Jenkins, PS, and ST mass
function fits with respect to the Warren fit for five different
redshifts over a range of halo masses.  Top to bottom: Redshifts $z=0$, 5,
10, 15, and 20.  Note that the ranges of the axes are different in the
different panels. We do not show the Jenkins fit below masses of
10$^{11} h^{-1} M_\odot$ at $z=0$, since it is not valid for such low masses
at that redshift.} 
\label{plotone}
\end{figure}

Jenkins combined high resolution
simulations for four different CDM cosmologies ($\tau$CDM, SCDM,
$\Lambda$CDM, and OCDM) spanning a mass range of over 3 orders of
magnitude ($\sim (10^{12}-10^{15}) \,h^{-1}M_{\sun}$), and including
several redshifts between $z=5$ and 0. Independent of the
underlying cosmology, the following fit provided a good representation
of their numerical results (within $\pm 20\%$):
\begin{equation}
f_{\rm Jenkins}(\sigma) = 0.315 \exp 
\left( -| \ln \sigma^{-1} + 0.61 |^{3.8} \right).
\end{equation}
The above formula is very close to the Sheth-Tormen fit, leading to
some improvement at the high-mass end. The disadvantage is that it
cannot be simply extrapolated beyond the range of the fit, since it was
tuned to a specific set of simulations.

By performing 16 nested-volume dark matter simulations, 
Warren was able to obtain significant halo
statistics spanning a mass range of 5 orders of magnitude~($\sim
(10^{10}-10^{15})\,h^{-1} M_{\sun}$).  Because this represents by
far the largest uniform set of simulations--based on multiple boxes
with the same cosmology run with the same code--we use it as a
reference standard throughout this paper. Using a functional form
similar to ST, Warren determined the best mass function fit
to be
\begin{equation}
f_{\rm Warren}(\sigma) = 0.7234 \left( \sigma^{-1.625} + 0.2538 \right) 
\exp \left( -\frac{1.1982}{\sigma^2}\right).
\end{equation}
For a quantitative comparison of the different fits at different
redshifts, we show the ratio of the PS, Jenkins, and ST fits with
respect to the Warren fit in Figure~\ref{plotone}.  We do not show the
Jenkins fit below $10^{11}\,h^{-1}M_\odot$ at $z=0$ since it diverges in
this regime. The original ST fit, the Jenkins fit, and the Warren fit
all give similar predictions.  The discrepancy between PS and the
other fits becomes more severe for higher masses at high redshifts.
PS dramatically underpredicts halos in the high-mass range at high
redshifts (assuming that the other fits lead to reasonable results in
this regime). For low-mass halos the disagreement becomes less
severe. For $z=0$ the Warren fit agrees, especially in the low-mass
range below $10^{13}\,h^{-1} M_\odot$, to better than 5\% with the ST
fit. At the high-mass end the difference increases up to 20\%. The
Jenkins fit leads to similar results over the considered mass range.
At higher redshifts and intermediate-mass ranges around
$10^{9}\,h^{-1}M_\odot$, the Warren and ST fit disagree by roughly a
factor of 2.

Several other groups have suggested modifications of the ST fit. In
\S\ref{resint} we compare our results with two of them.
\cite{Reed03} suggest an empirical adjustment to the ST fit by
multiplying it with an exponential function, leading to
\begin{equation}
f_{\rm Reed03}(\sigma)=f_{\rm ST}(\sigma)\,
\exp\left\{-0.7/\left[\sigma(\cosh(2\sigma))^5\right]\right\},
\end{equation}
valid over the range $-1.7\le \ln \sigma^{-1} \le 0.9$. This
adjustment leads to a suppression of the ST fit at large
$\sigma^{-1}$. In \cite{Reed07} the adjustment to the ST fit is
slightly modified again, leading to the following new fit:
\begin{eqnarray}
  f_{\rm Reed07}(\sigma)&=& A \sqrt{\frac{2a}{\pi}} \left[1 + 
\left(\frac{\sigma^2}{a\delta_c^2}\right)^p+0.6G_1+0.4G_2\right]\nonumber\\
&&\times\frac{\delta_c}{\sigma}
\exp\left[-\frac{ca\delta_c^2}{2\sigma^2} 
-\frac{0.03}{(n_{\rm eff}+3)^2}
\left(\frac{\delta_c}{\sigma}\right)^{0.6}\right],\\
G_1&=&\exp\left[-\frac{\ln(\sigma^{-1}-0.4)^2}{2(0.6)^2}\right],
\label{g1_def}\\
G_2&=&\exp\left[-\frac{\ln(\sigma^{-1}-0.75)^2}{2(0.2)^2}\right],
\label{g2_def}
\end{eqnarray} 
with $c=1.08$, $ca=0.764$, and $A=0.3222$. The adjustment has very
similar effects to that of \cite{Reed03}, as we show in
\S\ref{resint}. \cite{Reed07} note that the (small) suppression
of the mass function relative to ST as a function of redshift seen in
simulations (see also Heitmann et al. 2006a) can be treated by adding
an extra parameter, the power spectral slope at the scale of the halo
radius, $n_{\rm eff}$ (formally defined by equation~(\ref{n_eff_def})
below).  We return to this issue when we discuss our numerical
results in \S5.  We summarize the described, most commonly used 
fitting functions in Table~\ref{tabone}.

Although fitting functions may be a useful way to approximately
encapsulate results from simulations, meaningful comparisons to
observations require overcoming many hurdles, e.g., an operational
understanding of the definition of halo mass (see, e.g., White~2001),
how it relates to various observations, and error control in $N$-body
codes (see, e.g., O'Shea et al.~2005; Heitmann et al.~2005). In this
paper, our focus is first on identifying possible systematic problems
in the $N$-body simulations themselves and how they can be avoided and
controlled.

\subsection{Halo Growth Function}
\label{halog}

\begin{figure}[t]
  \plotone{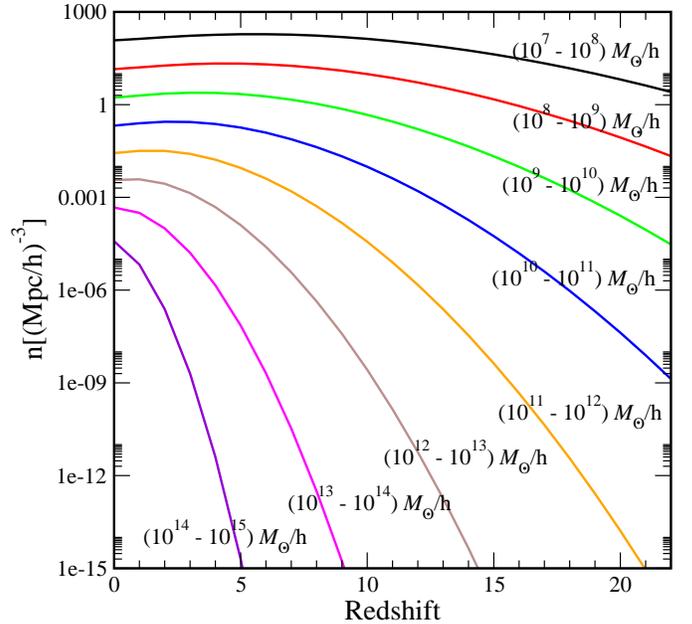}
\caption{Halo growth function based on the Warren mass function fit
  for different mass bins. The curves for the lower mass bins have a
  maximum at $z>0$ which reflects a crossover of the mass functions at
  different redshifts.}   
\label{plottwo}
\end{figure}

A useful way to study the statistical evolution of halo masses in
simulations is to transform the mass function into the halo growth
function, $n(M_1,M_2,z)\equiv \int_{M_1}^{M_2}F\,d\log M$ \cite[]{Heitmann06},
which measures the mass-binned number density of halos as a function
of redshift.  The halo growth function, plotted versus redshift in
Figure~\ref{plottwo}, shows at a glance how many halos in a particular
mass bin and box volume are expected to exist at a certain redshift.
This helps set the required mass and force resolution in a simulation
which aims to capture halos at high redshifts.  For a given simulation
volume, the halo growth function directly predicts the formation time
of the first halos in a given mass range.

In order to derive this quantity approximately, we
first convert an accurate mass function fit (we use the Warren fit
here) into a function of redshift $z$.  It has been shown recently by
us (Heitmann et al. 2006a) that mass function fits work reliably enough
out to at least $z=20$, and can therefore be used to estimate the halo
growth function. Figure~\ref{plottwo} shows the evolution of eight
different mass bins, covering the mass range investigated in this
paper, as a function of redshift $z$. As expected from the paradigm of
hierarchical structure formation in a $\Lambda$CDM cosmology, small
halos form much earlier than larger ones.  An interesting feature in
the lower mass bins is that they have a maximum at different
redshifts.  The number of the smallest halos grows until a redshift of
$~z=2$ and then declines when halos start merging and forming much more
massive halos.  This feature is reflected in a crossing of the mass
functions at different redshifts for small halos.

\subsection{Mass Function at High Redshift: Previous Work}
\label{evolreview}

\begin{figure}[t]
\begin{center}
\begin{center}
\leavevmode\includegraphics[width=7.5cm]{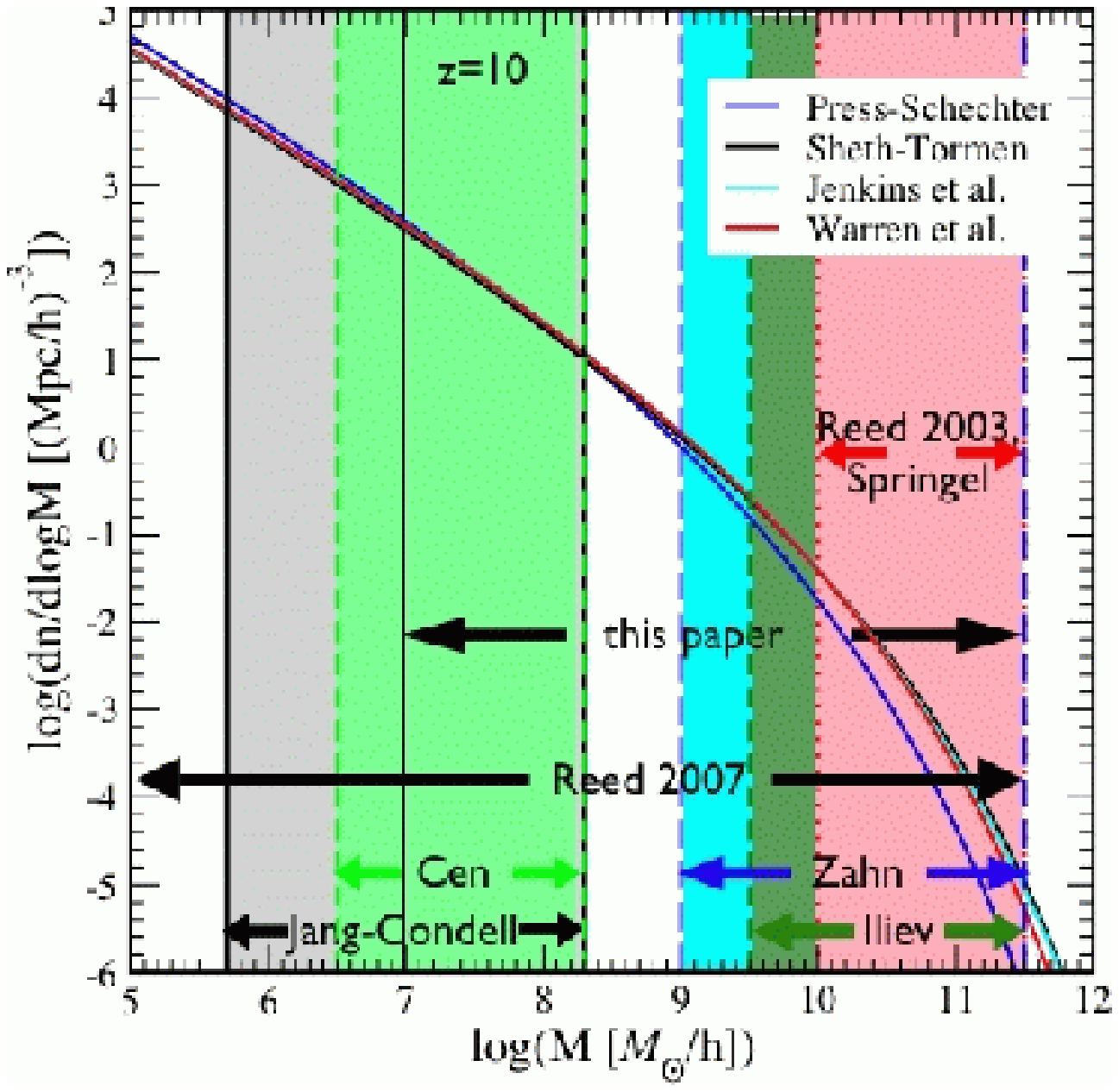}
\end{center}

\vspace{0.8cm}

\leavevmode\includegraphics[width=7.5cm]{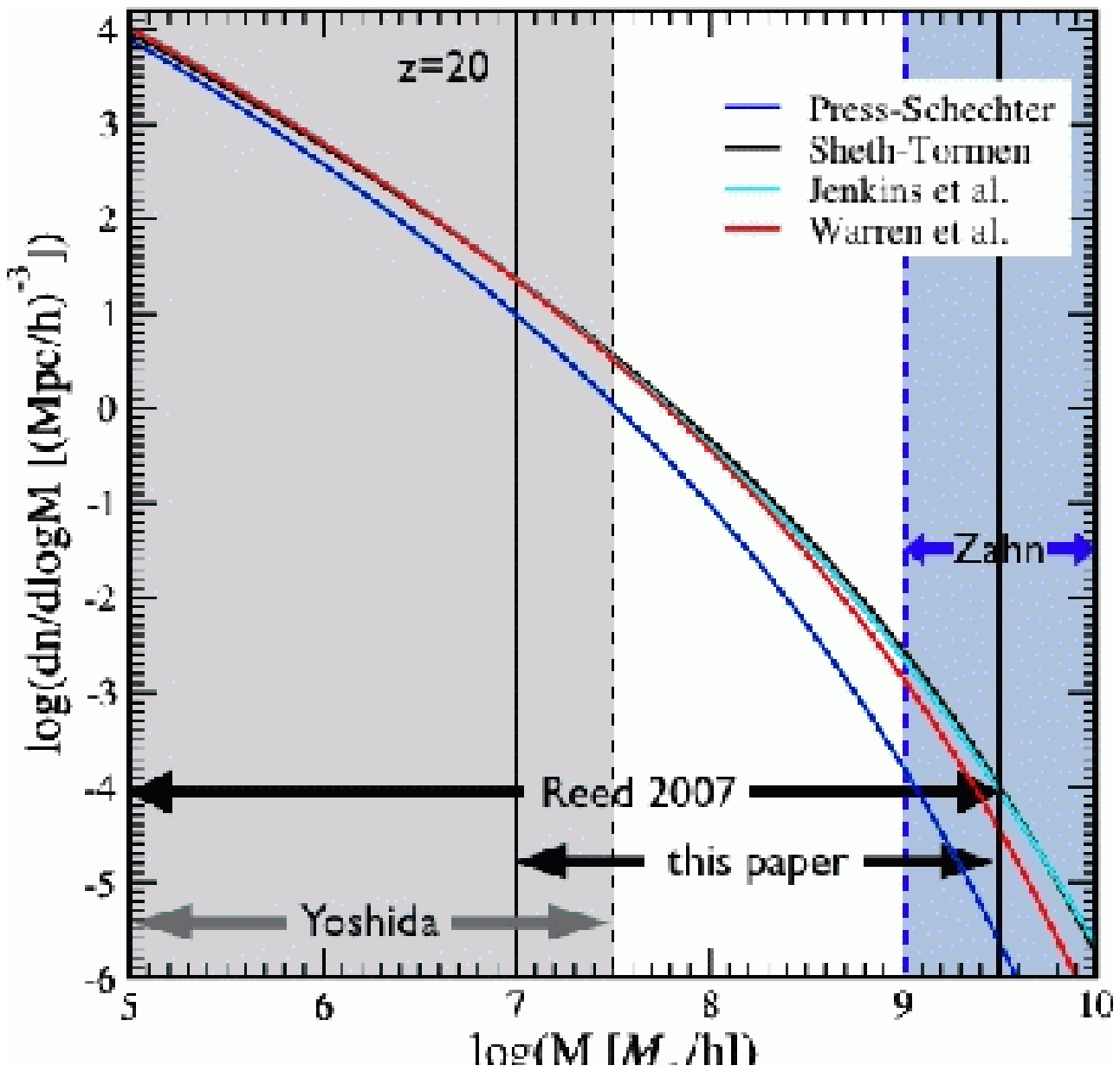}

\vspace{0.8cm}

\caption{Summary of recent work on the mass function at high redshift.
  The mass function fits are shown at $z=10$ (top) and $z=20$
  (bottom) for the cosmology used throughout this paper (the
  other groups used slightly different parameters). At $z=10$,
  Jang-Condell \& Hernquist (2001) (gray shaded region) cover the very
  low mass range using a very small box, as do Cen et al. (2004)
  (green shaded region).  The larger boxes of \cite{Reed07} and
  Springel et al. (2005) (red shaded region) lead to results at higher
  halo masses.  Note that in this regime the PS fit deviates
  substantially from the other fits, while at the very low mass end
  all fits tend to merge. Our suite of variable box sizes covers a
  mass range of 10$^7$ to 10$^{13.5}\,h^{-1}M_\odot$
  between $z=0$ and 20, a much larger range than previously
  covered by any group with a uniform set of simulations. At $z=20$
  Yoshida et al. (2003a, 2003b, 2003c, 2003d) 
  cover the very low mass end of the mass function,
  while Zahn et al. (2007) investigate larger mass halos. Our simulations
  overlap with both of them at the edges. By combining a heterogeneous
  set of simulations, \cite{Reed07} cover a wide range in mass and
  redshift. Figure quality reduced for the arXiv version of the paper.}
\label{plotthree}
\end{center}
\end{figure}

\begin{table*}
\begin{center}
\caption{\label{tabtwo} Summary of the Performed Runs}
\begin{tabular}{cccccccc}
\tableline\tableline
  & Box Size & Resolution & & &  Particle Mass & Smallest Halo & \\
\raisebox{1.4ex}[0pt]{Mesh} & ($h^{-1}$Mpc) & ($h^{-1}$kpc) &
\raisebox{1.4ex}[0pt]{$z_{\rm in}$} &
\raisebox{1.4ex}[0pt]{$z_{\rm final}$} & ($h^{-1}M_\odot$) & ($h^{-1}M_\odot$) &
\raisebox{1.4ex}[0pt]{No.~of Realizations} \\
\hline
 1024$^3$ & 256  & 250  & 100  & 0 & $8.35\times 10^{10}$ 
& $3.34 \times 10^{12}$  & 5\\
 1024$^3$ & 128  & 125  & 200  & 0 & $1.04\times 10^{10}$ 
& $4.18 \times 10^{11}$  & 5\\
 1024$^3$ &  64  & 62.5  & 200 & 0 & $1.31\times 10^9$   
& $5.22 \times 10^{10}$ &  5\\
 1024$^3$ &  32  & 31.25 & 150 & 5 & $1.63\times 10^8$   
& $6.52 \times 10^{9}$  & 5\\
 1024$^3$ &  16  & 15.63 & 200 & 5 &$2.04\times 10^7$  
& $8.16\times 10^{8}$  & 5\\
 1024$^3$ &   8  &  7.81 & 250 & 10 & $2.55\times 10^6$ 
& $1.02 \times 10^{8}$ & 20\\
 1024$^3$ &   4  &  3.91 & 500 & 10 & $3.19\times 10^5$ 
& $1.27 \times 10^{7}$ & 15\\
\tableline\tableline

\vspace{-1.5cm}

\tablecomments{Mass and force resolutions of the different runs. The
smallest halos we consider contain 40 particles. All simulations
have 256$^3$ particles.}
\end{tabular}
\end{center}
\end{table*}

Most of the effort to characterize, fit, and evaluate the mass
function from simulations has been focused on or near the current
cosmological epoch, $z\sim 0$.  This is mainly for two reasons: (1) so
far most observational constraints have been derived from low-redshift
objects ($z<1$); (2) the accurate numerical evaluation of the mass
function at high redshifts is a nontrivial task.

The increasing reach of telescopes on the ground and in space, such as
the upcoming James Webb Space Telescope, allows us to study the
Universe at higher and higher redshifts. Recent discoveries include
970 galaxies at redshifts between $z=1.5$ and $z=5$ from the VIMOS VLT
Deep Survey \cite[]{Lefevre05}, and the recent observation of a galaxy
at $z=6.5$ \cite[]{Mobasher05}. The epoch of reionization (EOR) is of
central importance to the formation of cosmic structure.  Although our
current observational knowledge of the EOR is rather limited, future
21 cm experiments have the potential for revolutionizing the
field. Proposed low-frequency radio telescopes include LOFAR (Low
Frequency Array)~\footnote{See http://www.lofar.org}, the Mileura Wide
Field Array
(MWA)~\cite[]{Bowman06}\footnote{See http://haystack.mit.edu/arrays/MWA/},
and the next-generation SKA (Square Kilometer
Array)~\footnote{See http://www.skatelescope.org}.  The observational
progress is an important driver for high-redshift mass function
studies.

Theoretical studies of the mass function at high redshifts are
challenging due to the small masses of the halos at early times.  In
order to capture these small-mass halos, high mass and force resolution
are both required. For the large simulation volumes typical in
cosmological studies, this necessitates a very large number of
particles, as well as very high force resolution. Such simulations are
very costly, and only a very limited number can be performed,
disallowing exploration of a wide range of possible simulation
parameters. Alternatively, many smaller volume simulation boxes, each
with moderate particle loading, can be employed. This leads
automatically to high force and mass resolution in grid codes (such as
particle-mesh [PM]) and also reduces the costs for achieving sufficient
resolution for particle codes (such as tree codes) or hybrid codes
(such as TreePM).  The disadvantages of this strategy are the limited
statistics in individual realizations (because fewer halos form in a
smaller box) and the unreliability of simulations below an
intermediate redshift at which the largest mode in the box is still
(accurately) linear.  In addition, results from small boxes may be
biased, since they only focus on a small region and volume. Therefore,
one must show that the simulations are free from finite-volume
artifacts, e.g.\ missing tidal forces, and run a sufficient number of
statistically independent simulations to reduce the sample variance.
Both strategies, employing large volume or multiple small-volume simulations,
have been followed in the past in order to obtain results at high
redshifts. The different mass ranges investigated by different groups
are shown in Figure~\ref{plotthree}. The fits are shown for redshifts
$z=10$ and 20. In the Appendix we provide a very detailed
discussion on previous findings as organized by simulation volume.

In summary, there is considerable variation in the high-redshift
($z>10$) mass function as found by different groups, independent of
box size and simulation algorithm. Broadly speaking, the results fall
into two classes: either consistent with linear theory scaling of a
universal form (Jenkins, Reed, ST, or Warren) at low redshift (Reed et
al. 2003, 2007; Springel et al. 2005; Heitmann et al. 2006a; Maio et
al. 2006; Zahn et al. 2007) or more consistent with
the PS fit (Jang-Condell \& Hernquist 2001; Yoshida et al. 2003a,
2003b, 2003c; Cen et al. 2004; Iliev et al. 2006; Trac \& Cen 2006).

Our aim here is to determine the evolution of the mass function
accurately, at the few percent level, and at the same time
characterize many of the numerical and physical factors that control
the error in the mass function (details below). We follow up on our
previous work~\cite[]{Heitmann06} and analyze a large suite of
$N$-body simulations with varying box sizes between 4 and
$256\,h^{-1}$Mpc, including many realizations of the small boxes, to
study the mass function at redshifts up to $z=20$ and to cover a large
mass range between $10^7$ and $10^{13.5}\,h^{-1}
M_\odot$. With respect to our previous work, the number of small-box
realizations has been increased to improve the statistics at high
redshifts. Our results categorically rule out the PS fit as being more
accurate than any of the more modern forms at {\em any} redshift up to
$z=20$, the discrepancy increasing with redshift.

\section{The Code and the Simulations}
\label{code}

All simulations in this paper are carried out with the parallel PM
code MC$^2$. This code solves the Vlasov-Poisson equations for an
expanding universe. It uses standard mass deposition and force
interpolation methods allowing periodic or open boundary conditions
with second-order (global) symplectic time stepping and fast fourier
transform based Poisson solves.  Particles are deposited on the
grid using the cloud-in-cell method.  The overall computational
scheme has proven to be accurate and efficient: relatively large
time steps are possible with exceptional energy conservation being
achieved.  MC$^2$ has been extensively tested against state-of-the-art
cosmological simulation codes (Heitmann et al.\ 2005, 2007).

We use the following cosmology for all simulations:
\begin{eqnarray}
&&\Omega_{\rm tot}=1.0,~~~\Omega_{\rm CDM}=0.253,~~~ 
\Omega_{\rm baryon}=0.048, \nonumber\\
&&\sigma_8=0.9,~~~ H_0=70\ {\rm km\ s^{-1}\,Mpc^{-1}},~~~n=1,
\end{eqnarray}
in concordance with cosmic microwave background and large scale
structure observations \cite[]{Mactavish05} (the third-year 
Wilkinson Microwave Anisotropy Probe 
observations suggest a lower value of $\sigma_8$;
\citet{Spergel06}). The transfer functions are generated with CMBFAST
\cite[]{Seljak96}.  We summarize the different runs, including their
force and mass resolution, in Table~\ref{tabtwo}.  As mentioned
earlier, we identify halos with a standard FOF halo finder with a
linking length of $b=0.2$.  Despite several shortcomings of the FOF
halo finder, e.g., the tendency to link up two halos which are close
to each other (see, e.g., Gelb \& Bertschinger~1994, Summers et
al.~1995) or statistical biases (Warren), the FOF algorithm
itself is well defined and very fast.  As discussed in
\S\ref{halomass}, we adopt the correction for sampling bias
given by Warren when presenting our results.

\section{Initial Conditions and Time Evolution}
\label{icevol}

In a near-ideal simulation with very high mass and force resolution,
the first halos would form very early.  By $z=50$, a redshift commonly
used to start cosmological simulations, a large number of small halos
would already be present (see, e.g., Reed et al.~[2005] for a discussion
of the first generation of star-forming halos).  In a more realistic
situation, however, the initial conditions at $z=50$ have of course no
halos, the particles having moved only the relatively small distance
assigned by the initial Zel'dovich step.  Only after the particles
have traveled a sufficient distance and come close together can they
interact locally to form the first halos. In the following we estimate
the redshift when the Zel'dovich grid distortion equals the
interparticle spacing, leading to the most conservative estimate for
the redshift of possible first halo formation.  From this estimate, we
derive the necessary criterion for the starting redshift for a given box
size and particle number.

\begin{figure}[t]
  \begin{center}
  \includegraphics[width=80mm]{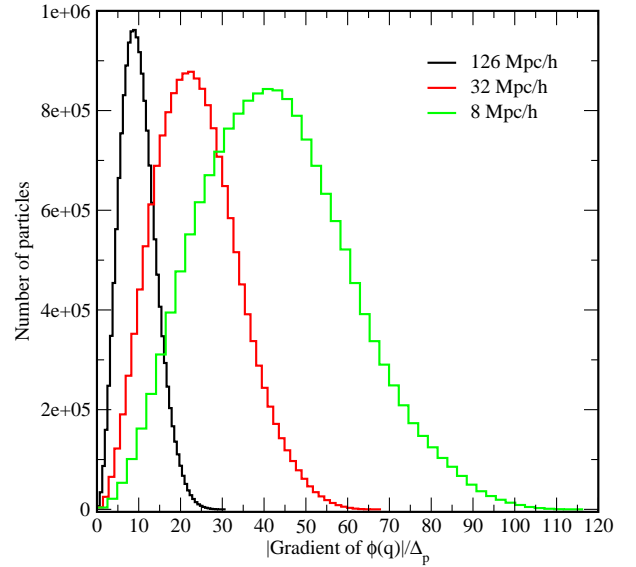}
  \end{center}

  \vspace{-0.2cm}

  \caption{Probability distribution of $|\nabla{\bf\phi}|$ in units of
    the interparticle spacing $\Delta_{\rm p}$.  All curves shown are drawn
    from $256^3$ particle simulations from an initial density grid of
    $256^3$ zones.  The physical box sizes are $126\,h^{-1}$Mpc (black
    line), $32\,h^{-1}$Mpc (red line), and $8\,h^{-1}$Mpc (green line).
    As expected, $\langle|\nabla{\bf\phi}|\rangle$ increases with
    decreasing box size (which is equivalent to increasing force
    resolution).  Therefore, $z_{\rm in}$ and $z_{\rm cross}$ are
    higher for the smaller boxes.}   
\label{plotfour}
\end{figure}

\subsection{Initial Redshift}
In order to capture halos at high redshifts, we have found that it is
very important to start the simulation sufficiently early. 
We consider two criteria for setting the starting redshift:
(1) ensuring the linearity of all the modes in the box used to sample the
initial matter power spectrum, and (2) restricting the initial particle
move to prevent interparticle crossing and to keep the particle grid
distortion relatively small. The first
criterion is commonly used to identify the starting redshift in
simulations. However, as shown below, it fails to provide sufficient
accuracy of the mass functions, accuracy which can be obtained when 
a second (much more restrictive) control is applied. 
Furthermore, it is important to allow a sufficient number of expansion 
factors between the starting redshift $z_{\rm in}$ and the highest 
redshift of physical significance. 
This is needed to make sure that artifacts
from the Zel'dovich approximation are negligible and that the memory
of the artificial particle distribution imposed at $z_{\rm in}$ (grid
or glass) is lost by the time any halo physics is to be extracted
from the simulation results. 

Although not studied here, it is important to note that high-redshift
starts do require the correct treatment of baryons as noted in
\S\ref{massdef}. In addition, redshift starts that are too high can lead
to force errors for a variety of reasons, e.g., interpolation
systematics, round-off, and correlated errors in tree codes.

\subsubsection{Initial Perturbation Amplitude}

\begin{table}[t]
\begin{center}
\caption{\label{tabthree} Initial Redshift Estimates from the Linearity of
  $\Delta^2(k_{\rm Ny})$}
\begin{tabular}{cccc}
\hline\hline
Box Size & $k_{\rm Ny}$ & & \\
($h^{-1}$Mpc) & ($h$\,Mpc$^{-1}$) &
\raisebox{1.4ex}[0pt]{$T(z=0,\, k_{\rm Ny})$} &
\raisebox{1.4ex}[0pt]{$z_{\rm in}$} \\
\hline
126 & 6.3  & 0.0002 & 33\\
32  & 25   & 1.7$\cdot 10^{-5}$ & 45\\
16  & 50   & 4.8$\cdot 10^{-6}$ & 50\\
8   & 100  & 1.3$\cdot 10^{-6}$ & 55\\
\hline\hline

\vspace{-1.5cm}

\tablecomments{The number of particles is $256^3$, the same in all 
simulations.}\end{tabular}
\end{center}
\end{table}

The initial redshift in simulations is often determined from the
requirement that all mode amplitudes in the box below the particle
Nyquist wavenumber characterized by $k_{\rm Ny}/2$ with $k_{\rm
  Ny}=2\pi/\Delta_{\rm p}$, where $\Delta_{\rm p}$ is the mean interparticle 
spacing, be sufficiently linear. The smaller the box
size chosen (keeping the number of particles fixed), the larger the
largest $k$-value.  Therefore, in order to ensure that the smallest
initial mode in the box is well in the linear regime, the starting
redshift must increase as the box size decreases.  In the following we
give an estimate based on this criterion for the initial redshift for
different simulation boxes.  We (conservatively) require the
dimensionless power spectrum $\Delta^2=k^3P(k)/2\pi^2$ to be smaller
than $0.01$ at the initial redshift. The initial power spectrum is
given by
\begin{equation}
\Delta^2(k_{\rm Ny},z_{\rm in})
=\frac{k^3 P(k_{\rm Ny},z_{\rm in})}{2\pi^2}
\sim\frac{B\ k^{n+3} T^2(k_{\rm Ny},z=0)}{2\pi^2 (z_{\rm in}+1)^2},
\end{equation}
where $B$ is the normalization of the primordial power spectrum (see,
e.g., Bunn \& White~[1997] for a fitting function for $B$ including
COBE results) and $T(k)$ is the transfer function.  We assume the
spectral index to be $n=1$, which is sufficient to obtain an estimate
for the initial redshift.  For a $\Lambda$CDM universe the
normalization is roughly $B\sim 3.4 \times 10^6 (h^{-1}{\rm Mpc})^4$.
Therefore, $z_{\rm in}$ is simply determined by
\begin{equation}
z_{\rm in}\simeq 4150 \ k_{\rm Ny}^2T(z=0,k_{\rm Ny}).
\end{equation}
We present some estimates for different box sizes in
Table~\ref{tabthree}.  For the smaller boxes ($<8~h^{-1}$Mpc), the
estimates for the initial redshifts are at around $z_{\rm in}=50$.

It is clear that this criterion simply sets a minimal requirement for
$z_{\rm in}$ and neglects the fact that the initial particle move should be small enough to maintain 
the dynamical accuracy of perturbation theory (linear or higher order) used to 
set the initial conditions. Also, 
this criterion certainly does not tell us that if, e.g., $z_{\rm
  in}=50$, then we may already trust the mass function at, say,
$z=30$. An example of this is provided by the results of \cite{Reed03},
who find that their high-redshift results between $z=7$ and 15 have not
converge if they start their simulations at $z_{\rm in}=69$. (A value
of $z_{\rm in}=139$ was claimed to be sufficient in their case.)

We now consider another criterion -- ostensibly similar in spirit --
that particles should not move more than a certain fraction of the
interparticle spacing in the initialization step. This second
criterion demands much higher redshift starts. 

\subsubsection{First Crossing Time}
\label{firstcross}

\begin{figure}[t]
  \begin{center}
  \includegraphics[width=80mm]{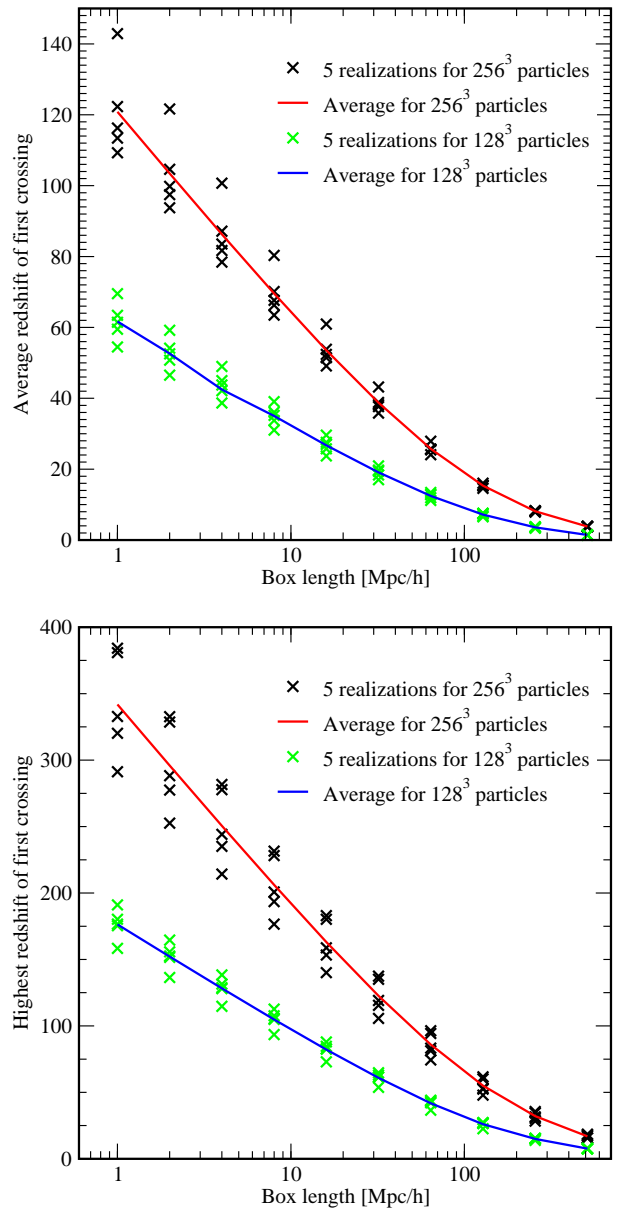}
  \end{center}

\caption{Average redshift of first crossing (top) and highest
  redshift of first crossing (bottom) as a function of box
  size. The initial conditions (five different realizations) are shown for
  boxes between 1 and 512$\,h^{-1}$Mpc with 128$^3$ and
  256$^3$ particles. For each initial condition, $z_{\rm cross}^{\rm
    first}$ and $z_{\rm cross}^{\rm rms}$ are shown by the
  crosses. The solid lines show the average from the five
  realizations. As expected, scatter from the different realizations is
  larger for smaller boxes. These plots provide estimates of the
  required initial redshift for a simulation since
  $|\nabla\phi|/\Delta_{\rm p}$ is $z$-independent in the Zel'dovich
  approximation (see text).}
\label{plotfive}
\end{figure}

In cosmological simulations, initial conditions are most often
generated using the Zel'dovich approximation \cite[]{Zeldovich70}.
Initially each particle is placed on a uniform grid or in a glass
configuration and is then given a displacement determined by the
relation
\begin{equation}
\label{zeldo}
{\bf x} = {\bf q} -d(z)\nabla{\bf\phi},
\end{equation}
where ${\bf q}$ is the Lagrangian coordinate of each particle.  The
gradient of the potential $\bf{\phi}$ is independent of the redshift
$z$.  The Zel'dovich approximation holds in the mildly nonlinear
regime, as long as particle trajectories do not cross each other (no
caustics have formed).  Studying the magnitude of $|{\nabla\bf \phi}|$
allows us to estimate two important redshift values:  first, the
initial redshift $z_{\rm in}$ at which the particles should not have
moved on average more than a fraction of the interparticle spacing
$\Delta_{\rm p}=L_{\rm box}/n_{\rm p}$, where $L_{\rm box}$ is the
physical box size and $n_{\rm p}$ the number of particles in the
simulation; second, the redshift at which particles first move more than
the interparticle spacing, $z_{\rm cross}$, i.e., 
at which they have traveled on
average a distance greater than $\Delta_{\rm p}$.

\begin{figure*}[t]
  \begin{center}
  \includegraphics[width=180mm]{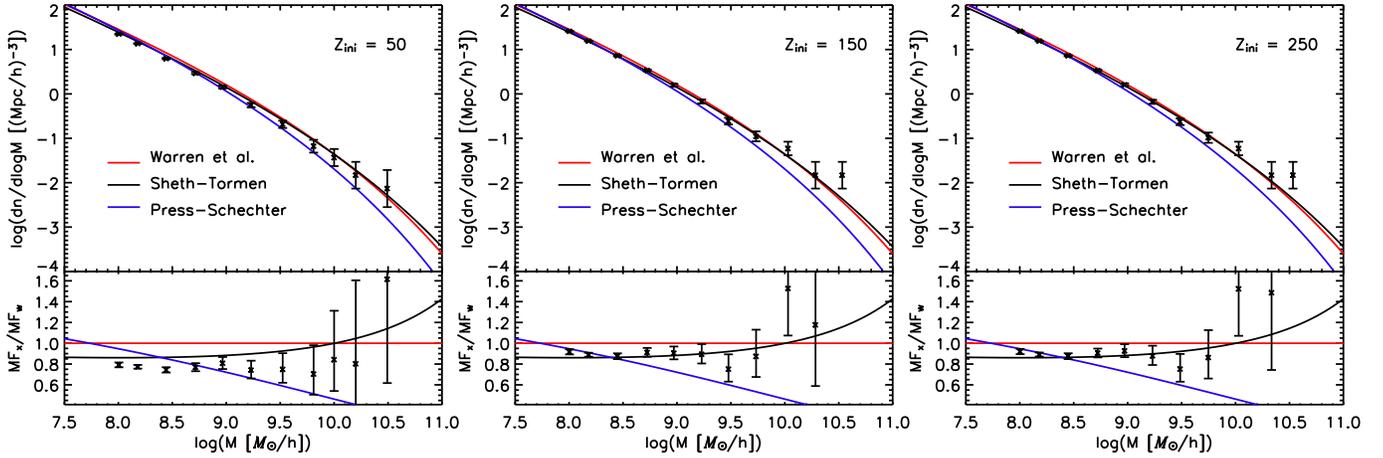}
  \end{center}

\caption{Dependence of the mass function on the initial redshift.
  The results are at $z=10$ from three 8$\,h^{-1}$Mpc box simulations with
  $z_{\rm in}=50$ (left), $z_{\rm in}=150$ (middle), and at $z_{\rm
    in}=250$ (right).  The mass function in the left panel is
  systematically lower than the other two by roughly 15\%. Poisson
  error bars are shown.}
\label{plotsix}
\end{figure*}

For a given realization of the power spectrum, the magnitude of
$|{\nabla\bf \phi}|$ depends on two parameters: the physical box size
and the interparticle spacing.  Together these parameters determine
the range of scales under consideration.  The smaller the box, the
smaller the scales; therefore, $|{\nabla\bf \phi}|$ increases and both
$z_{\rm in}$ and $z_{\rm cross}$ increase.  Increasing the resolution
has the same effect.  In Figure~\ref{plotfour} we show the probability
distribution function for $|{\nabla\phi}|$ for three different box
sizes, $8$, $32$, and $126\,h^{-1}$Mpc, representing values studied by
other groups, as well as in this paper.  To make the comparison between
the different box sizes more straightforward, we have scaled
$|{\nabla\bf \phi}|$ with respect to the interparticle spacing
$\Delta_{\rm p}$.  All curves are drawn from simulations with 256$^3$
particles on a 256$^3$ grid, in accordance with the set up of our
initial conditions.  The behavior of the probability function follows
our expectations: the smaller the box, or the higher the force
resolution, the larger the initial displacements of the particles on
average.  From the mean and maximum values of such a distribution we
can determine appropriate values for $z_{\rm in}$ and $z_{\rm cross}$.
For our estimates we assume $d(z)\simeq 1/(1+z)$, which is valid for
high redshifts.  The maximum and rms initial displacements of
the particles can then be easily calculated:
\begin{eqnarray}
\delta^{\rm max}_{\rm in}&\simeq&
\frac{{\rm max}(|\nabla\phi|/\Delta_{\rm p})}{1+z_{\rm in}},\\
\delta^{\rm rms}_{\rm in}&\simeq&
\frac{{\rm rms}(|\nabla\phi|/\Delta_{\rm p})}{1+z_{\rm in}}.
\end{eqnarray}
The very first ``grid crossing'' of a particle occurs when $\delta^{\rm
  max}_{\rm in}=1$; on average the particles have moved  
more than one particle spacing when $\delta^{\rm rms}_{\rm in}=1$.
This leads to the following estimates:
\begin{eqnarray}
z^{\rm first}_{\rm cross}&\simeq&
{\rm max}(\nabla\phi/\Delta_{\rm p})-1,\\
z^{\rm rms}_{\rm cross}&\simeq&
{\rm rms}(\nabla\phi/\Delta_{\rm p})-1.
\end{eqnarray}
We show these two redshifts in Figure~\ref{plotfive} for 10 different
box sizes ranging from 1 to 512$\,h^{-1}$Mpc and for
256$^3$ and 128$^3$ particles. The top panel shows the average
redshift of the first crossing as a function of box size (which
corresponds to the maximum in Fig.~\ref{plotfour}). The bottom panel
shows the redshift where the first ``grid crossing'' occurs
(corresponding to the right tail in Fig.~\ref{plotfour}).  To
estimate the scatter in the results, we have generated five different
realizations for each box. As expected, the small boxes show much more
scatter. The average redshift of the first crossing in the
1$\,h^{-1}$Mpc box varies between $z=63$ and 83, while there is
almost no scatter in the 512$\,h^{-1}$Mpc box. Since
$|\nabla\phi|/\Delta_{\rm p}$ is independent of redshift in the
Zel'dovich approximation, a simple scaling determines the appropriate
initial redshift from these plots. For example, if a particle should
not have moved more than 0.3$\Delta_{\rm p}$ on average at the initial
redshift, the average redshift of first crossing has to be multiplied
by a factor $1/0.3=3.\bar{3}$. For an $8\,h^{-1}$Mpc box this leads to
a minimum starting redshift of $z=230$, while for a $126\,h^{-1}$Mpc
box this suggests a starting redshift of $z_{\rm in}=50.$ The 128$^3$
particle curve can be scaled to the 256$^3$ particle curve by
multiplying by a factor of 2. Curves for different particle loadings
can be obtained similarly.

\vspace{0.2cm}

\subsection{Transients and Mixing}

The Zel'dovich approximation matches the exact density and velocity
fields to linear order in Lagrangian perturbation theory. Therefore,
there is in principle an error arising from the resulting discrepancy
with the density and velocity fields given by the exact growing mode
initialized in the far past.

This error is linear in the number of expansion factors between
$z_{\rm in}$ and the redshift of interest $z_{\rm phys}$. It has been
explored in the context of simulation error by \cite{Valageas02} and
by \cite{Crocce06}. Depending on the quantity being calculated, the
number of expansion factors between $z_{\rm in}$ and $z_{\rm phys}$
required to limit the error to some given value may or may not be easy
to estimate. For example, unlike quantities such as the skewness of
the density field, there is no analytical result for how this error
impacts the determination of the mass function. Neither does there
exist any independent means of validating the result aside from
convergence studies. Nevertheless, it is clear that to be
conservative, one should aim for a factor of $\sim 20$ in
expansion factor in order to anticipate errors at the several percent
level, a rule of thumb that has been followed by many $N$-body
practitioners (and often violated by others!). This rule of thumb
gives redshift starts that are roughly in agreement with the estimates
in the previous subsection. Convergence tests done for our simulations 
show that the suppression in the mass function is very small (less
than 1\%) for simulations whose evolution covers a factor of 15 in the 
expansion factor and can be up to 20\% for simulations that evolved by 
only 5 expansion factors. However, due to modest particle 
loads, we were unable 
to distinguish between the error induced by too few expansion factors
and the breakdown of the Zel'dovich approximation. 

Another possible problem, independent of the accuracy of the
Zel'dovich approximation, is the initial particle distribution
itself. Whether based on a grid or a glass, the small-distance
($k>k_{\rm Ny}$) mass distribution is clearly not sampled at all by
the initial condition. Therefore, unlike the situation that would
arise if a fully dynamically correct initial condition were given,
some time must elapse before the correct small-separation statistics
can be established in the simulation. Thus, all other things being
equal, for the correct mass function to exist in the box, one must run
the simulation forward by an amount sufficiently greater than the time
taken to establish the correct small-scale power on first-halo scales
while erasing memory on these scales of the initial conditions. If
this is not done, structure formation will be suppressed, leading to a
lowering of the halo mass function.

Because there is no fully satisfactory way to calculate $z_{\rm in}$ in
order to compute the mass function at a given accuracy, we subjected {\em
  every} simulation box to convergence tests in the mass function
while varying $z_{\rm in}$. The results shown in this paper are all
converged to the sub-percent level in the mass function. We give an
example of one such convergence test below.

\subsubsection{Initial Redshift Convergence Study}

As mentioned above, we have tested and validated our estimates for the
initial redshift for all the boxes used in the simulation suite via
convergence studies. Here, we show results for an 8$\,h^{-1}$Mpc box
with initial redshifts $z_{\rm in}=50$, 150, and 
250 in Figure~\ref{plotsix}, where the mass functions at $z=10$
are displayed.  For the lowest initial redshift, $z_{\rm in}=50$, the
average initial particle movement is 1.87$\Delta_{\rm p}$, while some
particles travel as much as 5.03$\Delta_{\rm p}$.  This clearly
violates the requirement that the initial particle grid distortion be
kept sufficiently below 1 grid cell.  The starting redshift $z_{\rm
  in}=150$ leads to an average displacement of 0.63$\Delta_{\rm p}$
and a maximum displacement of 1.71$\Delta_{\rm p}$, and therefore
just barely fulfills the requirements.  For $z_{\rm in}=250$ we find
an average displacement in this particular realization 
of 0.37$\Delta_{\rm p}$ and a maximum 
displacement of 1.00$\Delta_{\rm p}$.
 
The bottom plot in each of the three panels of Figure~\ref{plotsix}
shows the ratio of the mass functions with respect to the Warren fit.
In the middle and right panels the ratio for the largest halo is
outside the displayed range. The mass function from the simulation
started at $z_{\rm in}=50$ (left panel) is noticeably 
lower, $\sim 15\%$, than for the other two simulations.  The mass
functions from the two higher redshift starts are in good agreement,
showing that the choice for average grid distortion of 
approximately 0.3$\Delta_{\rm p}$ is 
conservative, and that one can safely use (0.5--0.6)$\Delta_{\rm p}$. 
The general 
conclusion illustrated by Figure~\ref{plotsix} is that if a simulation
is started too late, halos are found to be missing over the entire
mass range.  With the late start, there is less time to form bound
objects.  Also, some particles that are still streaming towards a halo
do not have enough time to join it. Both of these artifacts lead to an
overall downshift of the mass function.

To summarize, requiring a limit on initial displacements sets the
starting redshift much higher than simply demanding that all modes in
the box stay linear. Indeed, the commonly used latter criterion (with
$\delta^{\rm rms}\sim 0.1$) is not adequate for computing the halo
mass function at high redshifts. One must verify that the chosen
$z_{\rm in}$ sets an early enough start as shown here. We comment on
previous results from other groups with respect to this finding below
in \S\ref{conclusion}.

\subsection{Force and Mass Resolution}

We now take up an investigation of the mass and force resolution
requirements. The first useful piece of information is the size of the
simulation box: from Figure~\ref{plottwo} we can easily translate the
number density into when the first halo is expected to appear in a box
of volume $V$.  For example, a horizontal line at $n=10^{-6}$ would
tell us at what redshift we would expect on average to find 1 halo
of a certain mass in a $(100\,h^{-1}$Mpc)$^3$ box.  The first halo of
mass $~10^{11}-10^{12}\,h^{-1}M_\odot$ will appear at $z\simeq 15.5$,
and the first cluster-like object of mass
$~10^{14}-10^{15}\,h^{-1}M_\odot$ at $z\simeq 2$.  Of course, these
statements only hold if the mass and force resolution are sufficient to
resolve these halos.  The mass of a particle in a simulation, and
hence the halo mass, is determined by three parameters: the matter
content of the Universe $\Omega_{\rm m}$, including baryons and dark
matter, the physical box size $L_{\rm box}$, and the number of
simulation particles $n_{\rm p}^3$:
\begin{equation}
m_{\rm particle}= 2.775 \times 10^{11}\Omega_{\rm m} 
\left(\frac{L_{\rm box}}{n_{\rm p}\,h^{-1}{\rm Mpc}}\right)^3 \,h^{-1} {M}_\odot.
\label{mpart}
\end{equation}

\begin{figure*}[t]
  \begin{center}
  \includegraphics[width=180mm]{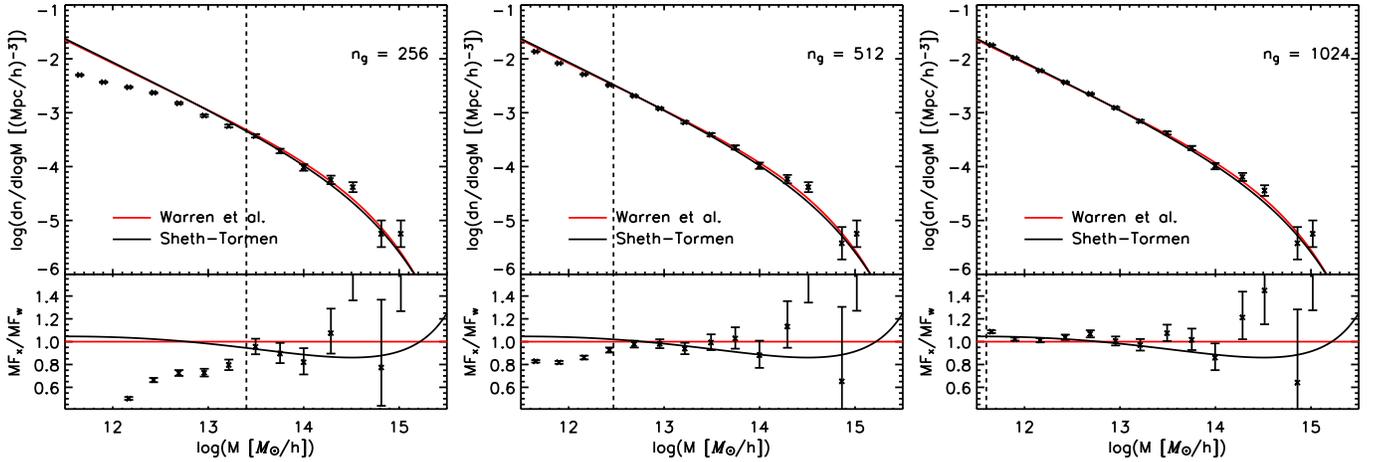}
  \end{center}
\caption{Convergence of the mass function as a function of force
  resolution. All results are shown at $z=0$, for 256$^3$ particles
  and a 126$\,h^{-1}$Mpc box with Poisson error bars.  The resolution
  varies between 256$^3$ (left), 512$^3$ (middle), and 1024$^3$ grid
  points (right). The vertical line denotes the predicted theoretical
  resolution limit: halos on the right of the line should not be
  lost. The resolution limit is 2500 particles per halo for the
  256$^3$ grid, 300 particles per halo for the $512^3$ grid, and 40
  particles per halo for the 1024$^3$ grid.}
\label{plotseven}
\end{figure*}

The required force resolution to resolve the chosen smallest halos can
be estimated very simply. Suppose we aim to resolve a virialized halo
with comoving radius $r_{\Delta}$ at a given redshift $z$, where
$\Delta$ is the overdensity parameter with respect to the critical
density $\rho_{\rm c}$. The comoving radius $r_\Delta$ is given by
\begin{equation}
r_\Delta=9.51 \times 10^{-5}
\left[\frac{\Omega(z)}{\Omega_{\rm m}}\right]^{1/3}\left(
\frac{1}{\Delta}\frac{M_{\Delta {\rm c}}}{h^{-1}{\rm M_\odot}}\right)^{1/3}
\,h^{-1}{\rm Mpc},
\label{rdelta}
\end{equation}
where $\Omega(z)=\Omega_{\rm m}(1+z)^3/[\Omega_{\rm
  m}(1+z)^3+\Omega_\Lambda]$ and the halo mass $M_{\Delta {\rm c}}=
m_{\rm part} n_{\rm h}$, where $n_{\rm h}$ is the number of
particles in the halo.  We measure the force resolution in terms of
\begin{equation}
\delta_{\rm f}=\frac{L_{\rm box}}{n_{\rm g}}.
\end{equation}
In the case of a grid code, $n_{\rm g}$ is literally the number of grid
points per linear dimension; for any other code, $n_{\rm g}$ stands for the
number of ``effective softening lengths'' per linear dimension. To
resolve halos of mass $M_{\Delta{\rm c}}$, a minimal requirement is
that the code resolution be smaller than the radius of the halo we
wish to resolve:
\begin{equation}
\delta_{\rm f} < r_\Delta. 
\label{resol}
\end{equation}
Note that this minimal resolution requirement is aimed only at
capturing halos of a certain mass, not at resolving their interior
profile. Next, inserting the expression for the particle
mass~(eq. [\ref{mpart}]) and the comoving radius~(eq. [\ref{rdelta}]) into the
requirement~(eq [\ref{resol}]) and employing the relation between the
interparticle spacing $\Delta_{\rm p}$ and the box size $\Delta_{\rm
  p}=L_{\rm box}/n_{\rm p}$, the resolution requirement reads
\begin{equation}
\frac{\delta_{\rm f}}{\Delta_{\rm p}}<
0.62\left[\frac{n_h\Omega(z)}{\Delta}\right]^{1/3}.
\label{resreq}
\end{equation}
We now illustrate the use of this simple relation with an example. Let
$\Delta=200$ and consider a $\Lambda$CDM cosmology with $\Omega_{\rm
  m}=0.3$. Then for PM codes for which $\delta_{\rm
  f}/\Delta_{\rm p}=n_{\rm p}/n_{\rm g}$, we have the following
conclusions.  If the number of mesh points is the same as the number
of particles ($n_{\rm p}=n_{\rm g}$), halos with less than 2500
particles cannot be accurately resolved.  If the number of mesh points
is increased to 8 times the particle number ($n_{\rm p}=1/2 n_{\rm
  g}$), commonly used for cosmological simulations with PM codes, the
smallest halo reliably resolved has roughly 300 particles, and if the
resolution is increased to a ratio of 1 particle per 64 grid cells,
which we use in the main PM simulations in this paper, halos with
roughly 40 particles can be resolved.  It has been shown
in~\cite{Heitmann05} that this ratio (1:64) does not cause collisional
effects and that it leads to consistent results in comparison to
high-resolution codes. Note that increasing the resolution beyond this
point will not help, since it is unreliable to sample halos with too
few particles. Note also that a similar conclusion holds for any
simulation algorithm and not just for PM codes.

In Figure~\ref{plotseven} we show results from a resolution convergence
test at $z=0$.  We run 256$^3$ particles in a 126$\,h^{-1}$Mpc box with
three different resolutions: 0.5, 0.25, and 0.125$\,h^{-1}$Mpc.  The
vertical line in each figure shows the mass below which the resolution
is insufficient to capture all halos following 
condition~(\ref{resreq}). In all three cases, the agreement with the
theoretical prediction is excellent.

\begin{figure}[b]
  \begin{center}
  \includegraphics[width=80mm]{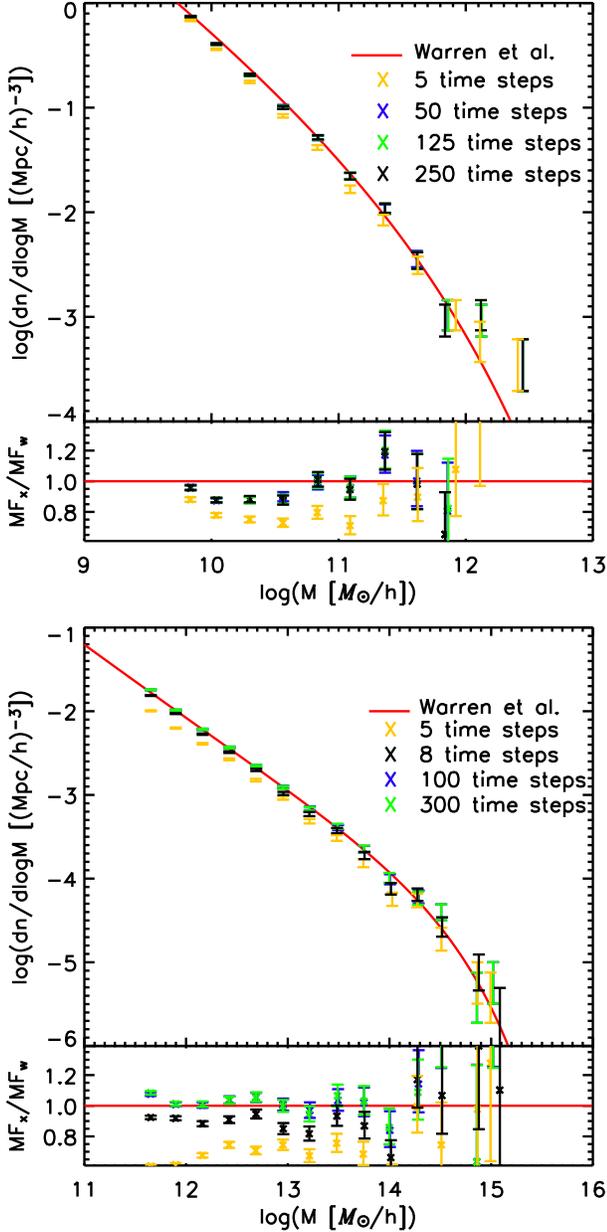}
  \end{center}
\caption{Top: One of the 32$\,h^{-1}$Mpc box realizations run
  with 250, 125, 50 and 5 time steps between $z_{\rm in}=150$ and
  $z_{\rm final}=5$. The mass function is shown at the final redshift
  $z=5$. Data points for all runs except the one with five time steps are
  so close that they are difficult to distinguish.  Bottom: 
  A 126$\,h^{-1}$Mpc box with 300, 100, 8, and 5 time steps between $z_{\rm
    in}=50$ and $z_{\rm final}=0$. The agreement for the very large
  halos for 100 and 300 time steps is essentially perfect. Poisson
  error bars are shown.}
\label{ploteight}
\end{figure}

\subsection{Time Stepping}

Next, we consider the question of time-step size and estimate the
minimal number of time steps required to resolve the halos of interest. We
begin with a rough estimate of the characteristic particle velocities
in halos. For massive halos, the halo mass $M_{200}$ and its velocity
dispersion are connected by the approximate relation (Evrard 2004):
\begin{equation}
M_{200}\simeq\frac{10^{15}\,h^{-1}M_\odot}{H/H_0}
\left(\frac{\sigma_{v}}{1080
{\rm \,km/s}}\right)^3.
\end{equation}
A more accurate expression can be found in \cite{Evrard07}, but the
above is more than sufficient for our purposes. At high redshift,
$\Omega_\Lambda$ can be neglected, and we can express the velocity
dispersion as a function of redshift:
\begin{equation}
\sigma_{\rm v}\simeq
10^{-2}\sqrt{1+z}\left(\frac{M_{200}}{\,h^{-1}M_\odot}\right)^{1/3}{\rm km s^{-1}}.
\end{equation}
In a time $\delta t$, the characteristic scale length $\delta l$ is
given by $\delta l\simeq \sigma_{\rm v}\delta t$ or
\begin{equation}
\label{deltat}
\delta t\simeq\frac{\delta l}{\sigma_{v}}=\frac{100\,\delta l/{\rm km}}{\sqrt{1+z}}
\left(\frac{M_{200}}{\,h^{-1}M_\odot}\right)^{-1/3} {\rm s}.
\end{equation}
The scale factor $a$ is a convenient time variable for codes working
in comoving units, such as ours. Expressed in terms of the scale
factor, equation~(\ref{deltat}) reads:
\begin{equation}
\delta a \simeq  10^4\frac{\delta l}{h^{-1}{\rm Mpc}}\left( \frac{M_{200}}
{h^{-1}M_\odot}\right)^{-1/3}.
\end{equation}
We are interested in the situation where $\delta l$ is actually the
force resolution, $\delta_{\rm f}$. In a single time step, the
distance moved should be small compared to $\delta_{\rm f}$; i.e., the
actual time step should be smaller than $\delta a$ estimated from the
above equation when $\delta l$ is replaced on the right--hand side with
$\delta_{\rm f}$. Let us consider a concrete example for the case of a
PM code where $\delta_{\rm f}=L_{\rm box}/n_{\rm g}$ as explained
earlier. For a ``medium'' box size of $L_{\rm box}=256\,h^{-1}{\rm Mpc}$
and a grid size of $n_{\rm g}=1024$, $\delta_{\rm f}=0.25\,h^{-1}{\rm
  Mpc}$. For a given box, the highest mass halos present have the
largest $\sigma_{\rm v}$ and give the tightest constraints on the time
step. For the chosen box size, a good candidate halo mass scale is
$M_{200}\sim 10^{15}\,h^{-1} M_\odot$ (this could easily be less, but it
does not change the result much). In this case,
\begin{equation}
\delta a \simeq 0.025.
\end{equation}
If, for illustration, we start a simulation at $z=50$ and evolve it
down to $z=0$, this translates to roughly 40 time steps.  We stress that
this estimate is aimed only at avoiding disruption of the halos
themselves, and is certainly not sufficient to resolve the {\it inner}
structure of the halo.

In Figure~\ref{ploteight} we show two tests of the time step criterion.
The top panel shows the result from a 32$\,h^{-1}$Mpc box at redshift
$z=5$. The simulation starts at $z_{\rm in}=150$ and is evolved with 50, 125,
and 250 time steps down to $z=5$. Following the argument above for this
box size, one would expect all three choices to be acceptable, and the
excellent agreement across these runs testifies that this is indeed
the case. We also carried out a run with only five time steps, which
yields a clearly lower ($\sim 20 \%$) mass function than the
others, but not as much as one would probably expect from such an
imprecise simulation.

The bottom panel shows the results from a 126$\,h^{-1}$Mpc box at
$z=0$. This simulation was started at $z{\rm in}=50$ and run to $z=0$
with 5, 8, 100, and 300 time steps. Again, as we would predict, the
agreement is very good for the last two simulations, and the
convergence is very fast, confirming our estimate that only
${\cal{O}}(10)$ time steps is enough to get the correct halo mass
function.  Overall, the halo mass function appears to be a very robust
measure, not very sensitive to the number of time steps. Nevertheless,
we used a conservatively large number of time steps, e.g., 500 for the
simulations stopping at $z=0$ and 300 for those stopping at $z=10$.

In the previous subsections we have discussed and tested different
error control criteria for obtaining the correct simulated mass
function at all redshifts. These criteria are (1) a sufficiently
early starting redshift to guarantee the accuracy of the Zel'dovich
approximation at that redshift and provide enough time for the halos
to form; (2) sufficient force and mass resolution to resolve the halos
of interest at any given redshift; and (3) sufficient numbers of time
steps. Violating any of these criteria {\em always} leads to a
suppression of the mass function. Most significantly, our tests show
that a late start (i.e., starting redshift too low) leads to a
suppression over the entire mass range under consideration, and is a
likely explanation of the low mass function results in the literature.
As intuitively expected, insufficient force resolution leads to a
suppression of the mass function at the low-mass end, while errors
associated with time stepping are clearly subdominant and should not
be an issue in the vast majority of simulations.

\section{Results and Interpretation}
\label{resint}

In this section we present the results from our simulation suite. We
describe how the data are obtained as well as the 
post-processing corrections applied. The latter include compensation
for FOF halo mass bias induced by finite (particle number) sampling,
and the (small) systematic suppression of the mass function induced by
the finite volume of the simulation boxes.

\subsection{Binning of Simulation Data}

Before venturing into the simulation results, we first describe how
they were obtained and reported from individual simulations.
We used narrow mass bins while conservatively keeping the
statistical shot noise of the binned points no worse than some given
value.  Bin widths $\Delta \log M$ were chosen such that the bins
contain an equal number of halos $N_{\rm h}$. The worst-case situation
occurs at $z=20$ for the 8$\,h^{-1}$Mpc box, which has $N_{\rm h}=80$;
the 4$\,h^{-1}$Mpc box at the same redshift has $N_{\rm h}=400$. At
$z=15$ we have $N_{\rm h}=150$, 1600, and 3000 for box sizes 16, 8, and
4$\,h^{-1}$Mpc, respectively. At $z=10$ the smallest value $N_{\rm
  h}=450$ is for the 32$\,h^{-1}$Mpc box, while at $z=5$ and 0 we
essentially always have $N_{\rm h}>10000$.

With a mass function decreasing monotonically with $M$, this binning
strategy results in bin widths increasing monotonically with $M$. The
increasing bin size may cause a systematic deviation -- growing
towards larger masses -- from an underlying ``true'' continuous mass
function. The data points for the binned mass function give the
average number of halos per volume in a bin,
\begin{equation}
\bar{\mf}\equiv N_{\rm h}/(V\Delta\log M),
\label{fbar}
\end{equation} 
plotted versus an average halo mass, averaged by the {\em number of
  halos} in the bin: 
\begin{equation}
\bar{M}\equiv \sum_{\rm bin}M/N_{\rm h}. 
\label{mbar}
\end{equation}
Assuming that the true mass function $dn/d\log M$ has some
analytic form $\mf(M)$, a systematic deviation
due to the binning prescription 
\begin{equation}
\epsilon_{\rm{bin}}\equiv\frac{\bar{\mf}-\mf(\bar{M})}{\mf(\bar{M})}
\end{equation}
can be evaluated by computing
$\bar{\mf}$ and $\bar{M}$ as
\begin{equation}
\bar{\mf}=\frac{\int_{\Delta M} dn}{\Delta\log M},~~~\bar{M}
       =\frac{\int_{\Delta M} Mdn}{\int_{\Delta M}dn}, 
\end{equation}
where $dn\equiv F(M)\,d\log M$ and the integrations are over a mass range
$[M,M+\Delta M]$. For the leading-order term of the Taylor expansion of
$\epsilon_{\rm bin}(\Delta M)$, we find
\begin{equation}
\epsilon_{\rm{bin}}\simeq\frac{F^{\prime\prime}-2(F^{\prime})^2/F}{24F}(\Delta M)^2,
\label{epsb}
\end{equation}
where the primes denote $\partial/\partial M$. A characteristic
magnitude of this $\epsilon_{\rm bin}$ for a general $\mf(M)$ is
$(\Delta M/M)^2/24$.  However, in our case, where the relevant scales
$k\gg k_{\rm eq}\sim0.01 h$Mpc$^{-1}$, $\epsilon_{\rm bin}$~has a much
stronger suppression, as explained below.

We know that the mass function is close to the universal form,
\begin{equation}
\mf(M)= \frac{\rho_{\rm b}}{M} f(\sigma)\frac{d\ln\sigma^{-1}}{d\log M}
\label{univ2}
\end{equation}
(see, eq.~[\ref{fsigma}]). Note that for $k\gg k_{\rm eq}$,
$\sigma^{-1}(M)$ is a slowly varying function, i.e.,
\begin{equation}
\frac{d\log \sigma^{-1}}{d\log M} \equiv \frac{n_{\rm eff}+3}{6}
\label{n_eff_def}
\end{equation}
is much smaller than unity, and the derivative $d\log
\sigma^{-1}/d\log M$ also changes slowly with~$M$.  Then, despite the
steepness of $\mf(\sigma)$ at small $\sigma$, the factor
$f(\sigma)\,d\ln\sigma^{-1}/d\log M$ in equation~(\ref{univ2})
depends weakly on $M$. Therefore, the mass function $\mf(M)$ is close
to being inversely proportional to $M$. In the limit of exact inverse
proportionality, $\mf\propto M^{-1}$, equation~(\ref{epsb}) tells us
that $\epsilon_{\rm{bin}}\rightarrow 0$. This effective cancellation
of the two terms on the right-hand side of equation~(\ref{epsb}) makes
the binning error negligible to the accuracy of our $\mf(M)$
reconstruction whenever a bin width $\Delta \log M$ does not exceed
$0.5$. To confirm the absence of any systematic offsets due to the
binning, we binned the data into $\log M$
intervals 5 times narrower and wider, 
with no apparent change in the inferred $\mf(M)$ dependence.

We remark that the situation could be quite different with another
binning choice. For example, if the binned masses $\bar{M}$ were
chosen at the centers of the corresponding $\log M$ intervals, $\log
\bar M=[\log M+\log(M+\Delta M)]/2$, the systematic binning deviation
\begin{equation}
\epsilon_{\rm{bin}}^{({\rm center})}\simeq
\frac{F^{\prime\prime}+F^{\prime}/M}{24F}(\Delta M)^2
\end{equation}
would have no special cancellation for the studied type of mass function.
A corresponding binning error would be about 2 orders of magnitude
larger than that of equations~(\ref{fbar}) and (\ref{mbar}).

The statistical error bars used are Poisson errors, following the
improved definition of Heinrich~(2003):
\begin{equation}
\label{heinrich}
\sigma_{\pm}=\sqrt{N_{\rm h}+\frac{1}{4}}\pm \frac{1}{2}.
\end{equation}
At large values of $N_{h}$, these error bars asymptote to the familiar
form~$\sqrt N_{\rm h}$. At smaller values of $N_{\rm h}$ -- which are of
minor concern here -- equation \ref{heinrich} has several advantages over the standard
Poisson error definition, some being (1) it is nonzero for $N_{\rm
  h}=0$; (2) the lower edge of the error bar does not go all the way
to zero when $N_{\rm h}=1$; (3) the asymmetry of the error bars
reflects the asymmetry of the Poisson distribution.

Finally, as noted earlier and discussed in the next section, all the
results shown in the following include a correction for the sampling
bias of FOF halos according to equation~(\ref{halocorr}). This mass
correction brings down the low-mass end of the mass function.

\subsection{FOF Mass Correction}
\label{masscorr}

\begin{figure}[t]
\begin{center}
\leavevmode\includegraphics[width=7.5cm]{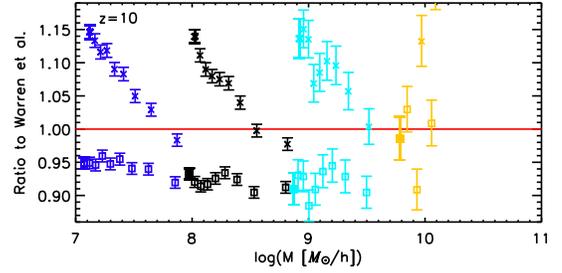}
\end{center}
\caption{FOF mass correction for halos in 4 (dark blue), 8 (black), 16
  (light blue), and 32 (yellow) $h^{-1}$Mpc boxes. To show the effect
  clearly, we plot the ratio of our data to the Warren fit. 
  Crosses show the uncorrected mass function and squares the mass function
  after correction, following eq.~(\ref{halocorr}). Note the
  smooth behavior of the corrected mass function as opposed to the
  mass-function jumps across box sizes for the uncorrected data.}
\label{plotnine}
\end{figure}

The mass of a halo as determined by the FOF algorithm displays a
systematic bias with the number of particles used to sample the
halo. Too few particles lead to an increase in the estimated halo
mass. By systematically subsampling a large halo population from
N-body simulations (at $z=0$), Warren determined an empirical
correction for this undersampling bias. For a halo with $n_{\rm h}$
particles, his correction factor for the FOF mass is given by
\begin{equation}
\label{halocorr}
n_{\rm h}^{\rm corr}=n_{\rm h}\left(1-n_{\rm h}^{-0.6}\right).
\end{equation}
We have carried out an independent exercise to check the systematic
bias of the FOF halo mass as a function of particle number based on
Monte Carlo sampling of an NFW halo mass profile with varying
concentration and particle number, as well as by direct checks against
simulations (e.g., Fig.~\ref{plotnine}); our results are
broadly consistent with equation~(\ref{halocorr}). Details will be
presented elsewhere (Z. Luki\'c et al., in preparation). In this
associated work we also address how overdensity masses connect to FOF
masses, how this relation depends on the different linking length used
for the FOF finder, and the properties of the halo itself, such as
the concentration.

The effect of the FOF sampling correction can be quickly gauged by
considering a few examples: for a halo with 50 particles, the mass
reduction is almost 10\%, for a halo with 500 particles, it is $\sim
2.4\%$, and for a well-sampled halo with 5000 particles, it is only
0.6\%. As a cautionary remark, this correction formula does not
represent a general recipe but can depend on variables such as the
halo concentration. Since the conditions under which different
simulations are carried out can differ widely, corrections of this
type should be checked for applicability on a case-by-case basis. Note
also that the correction for the mass function itself depends on how
halos move across mass bins once the FOF correction is taken into
account.

The choice of the mass function range in a given simulation box always
involves a compromise: too wide a dynamic range leads to poor
statistics at the high-mass end and possible volume-dependent
systematic errors, and too narrow a range leads to possible undersampling
biases. Our choice here reflects the desire to keep good statistical
control over each mass bin at the expense of wide mass coverage,
compensating for this by using multiple box sizes. Therefore, in our
case it is important to demonstrate control over the FOF mass
bias. An example of this is shown in Figure~\ref{plotnine}, where
results from four box sizes demonstrate the successful application of
the Warren correction to simulation results at $z=10$.

\subsection{Simulation Mass and Growth Function}
\label{simvol}

The complete set of simulations, summarized in Table~\ref{tabtwo},
allows us to study the mass function spanning the redshift range from
$z=20$ to 0. The mass range covers dwarf to massive galaxy halos
at $z=0$ (cluster scales are best covered by much bigger boxes as in
Warren and \citealt{Reed07}), and at higher redshifts goes
down to $10^7\,h^{-1}M_\odot$, the mass scale above which gas
in halos can cool via atomic line cooling (\citealt{Tegmark97}).

\subsection{Time Evolution of the Mass Function}
\label{timeevo}

\begin{figure*}
\begin{center}
\leavevmode\includegraphics[width=160mm]{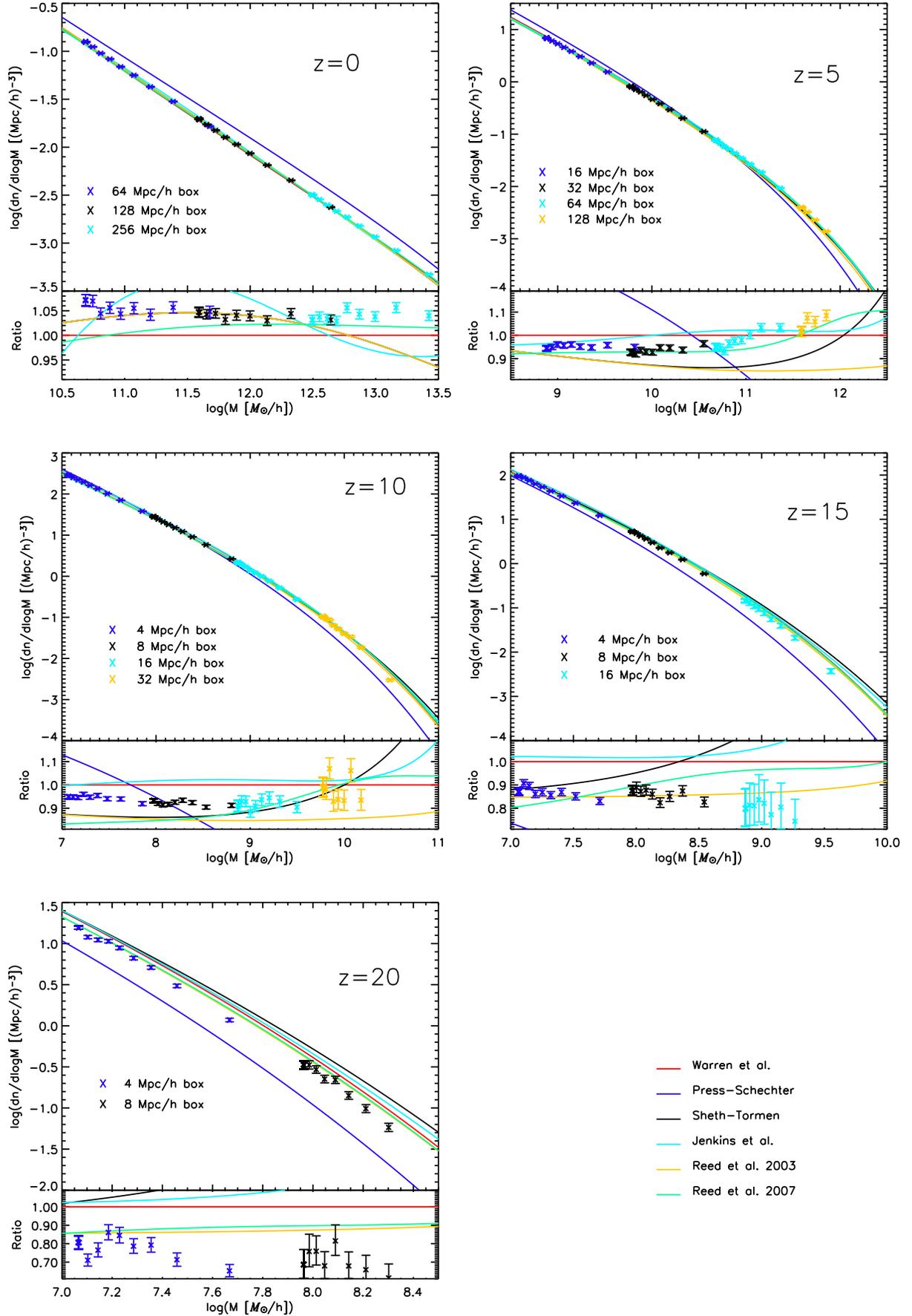}
\end{center}
\caption{Mass function at five different redshifts ($z=0$, 5, 10, 15,
  and 20; top to bottom) compared to different fitting formulae. 
  Note that the mass ranges are different at different
  redshifts.
  The
  simulation results have been corrected for FOF bias following
  Warren but not for finite-volume effects (for these,
  see Fig.~12). The bottom panel shows the ratio with respect to the
  Warren fit. Our simulations agree with the Warren fit at the 10\%
  level for redshifts smaller than 10, although there is a systematic
  offset of 5\% at $z=0$, where our numerical results are higher than
  the fit. At higher redshifts, the agreement is still very good (at
  the 20\% level) and becomes very close once finite-volume
  corrections are applied (Fig.~\ref{plottwelve}). PS is a bad fit
  at all redshifts, and especially at high redshifts, where the
  difference between PS and the simulation results is an order of
  magnitude.}
\label{plotten}
\end{figure*}

Halo mass functions from the multiple-box simulations are shown in
Figure~\ref{plotten}, with results being reported at five different
redshifts with no volume corrections applied. The combination of box
sizes is necessary because larger boxes do not have the mass
resolution to resolve very small halos at early redshifts, while
smaller boxes cannot be run to low redshifts. The bottom plot of each
panel shows the ratio of the numerically obtained mass function,
and various other fits, to the Warren fit as scaled by linear
theory (for volume-corrected results, see
Fig.~\ref{plottwelve}). Displaying the ratio has the advantage over
showing relative residuals that large discrepancies (more than 100\%)
appear more clearly. For all redshifts, the agreement with the Warren
fit is at the 20\% level. The ST fit matches the simulations for small
masses very well but overpredicts the number of halos at large
masses. This overprediction becomes worse at higher redshifts. For
example, at $z=15$ ST overpredicts halos of 10$^9 \,h^{-1}M_\odot$ by
a factor of 2. Reed et al. (2003) found a similar result: the ST fit
at $z=15$ for halos with mass larger than $10^{10} \,h^{-1}M_\odot$
disagrees with their simulation by 50\%. Agreement with the
Reed et al.~(2003, 2007) fits is also good, within the 10\%
level. (For a further discussion focused around the question of
universality, see Section~5.7.) The PS fit in general is not
satisfactory over a larger mass range at any redshift. It crosses the
other fits at different redshifts for different masses. Away from this
crossing region, however, the disagreement can be as large as an order
of magnitude, e.g. for $z=20$ over the entire mass range we consider
here.

\subsection{Halo Growth Function}
\label{halogres}

\begin{figure}[h]
  \includegraphics[width=80mm]{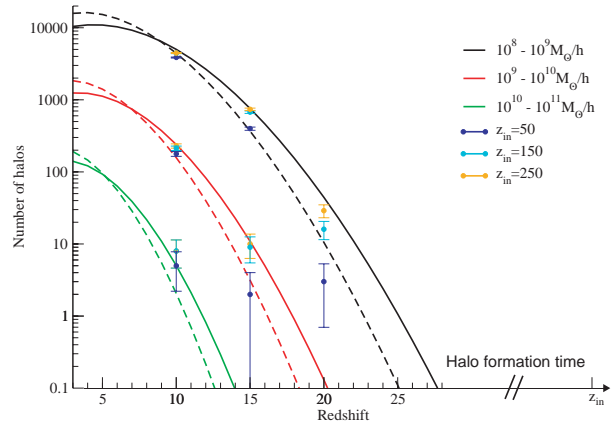}
\caption{Halo growth function for an 8$\,h^{-1}$Mpc box started from 
three different redshifts. The blue data points results from the 
$z=50$ start, the turquoise data points from the $z=150$ start, and 
orange from the $z=250$ start, which is the redshift satisfying our 
starting criteria. The two fits shown are the Warren fit (solid line) 
and the PS fit (dashed line). Three different mass bins are shown. It 
is interesting to note that the late start seems to follow the PS fit 
at high redshift.}
\label{ploteleven}
\end{figure}

As discussed in \S\ref{halog} the halo growth function (the
number density of halos in mass bins as a function of redshift) offers
an alternative avenue to study the time evolution of the mass
function. Figure~\ref{ploteleven} shows the halo growth function for
an 8$\,h^{-1}$Mpc box for three different starting redshifts, $z_{\rm
  in}=50$, 150, and 250 (these are the same simulations as in
Fig.~\ref{plotsix}).  The results are displayed at three redshifts,
$z=20$, 15, and 10 and for three mass bins, $10^8 -
10^9\,h^{-1}M_\odot$, $10^9 - 10^{10}\,h^{-1}M_\odot$, and $10^{10} -
10^{11}\,h^{-1}M_\odot$.

Assuming that the Warren fit scales at least approximately to high
redshifts, the first halos in the lowest mass bin are predicted to
form at $z_{\rm form}\sim 25$ (see Fig.~\ref{plotsix}). We have
found that if $z_{\rm form}$ is not sufficiently far removed from
$z_{\rm in}$, formation of the first halos is significantly
delayed/suppressed. In turn, this leads to suppressions of the halo
growth function and the mass function at high redshifts. As shown in
Figure~\ref{ploteleven}, the suppression can be quite severe at high
redshifts: the simulation result at $z=20$ from the late start at
$z_{\rm in}=50$ is an {\em order of magnitude} lower than that from
$z_{\rm in}=250$.  At lower redshifts, the discrepancy decreases, and
results from late-start simulations begin to catch up with the results
from earlier starts. Coincidentally, the suppression due to the late
start at $z_{\rm in}=50$ is rather close to the PS prediction which is
very significantly below the Warren fit in the mass and redshift range
of interest (see Fig.~\ref{ploteleven}). We take up this
point further below.

\subsection{Finite-Volume Corrections}

The finite size of simulation boxes can compromise results for the
mass function in multiple ways.  It is important to keep in mind that
finite-volume boxes cannot be run to lower than some redshift, $z_{\rm
  final}$, the stopping point being determined by when nonlinear
scales approach close enough to the box size. Approaching too near 
this point delays the ride-up of nonlinear power towards the low-$k$
end, with a possible suppression of the mass function. 

As a consequence of this delay, the evolution (incorrectly) appears
more linear at large scales than it actually should, as compared to
the $P(k)$ obtained in a much bigger box. Therefore, verifying linear
evolution of the lowest $k$-mode is by itself {\em not} sufficient to
establish that the box volume chosen was sufficiently large. For all
of our overlapping-volume simulations we have checked that the power
spectra were consistent across boxes up to the lowest redshift from
which results have been reported (Table~1 lists the stopping
redshifts).

Aside from testing for numerical convergence, it is important to show
that finite-volume effects are also under control, especially any
suppression of the mass function with decreasing box size (due to lack
of large-scale power on scales greater than the box size). Several
heuristic analyses of this effect have appeared in the
literature. Rather than rely solely on the unknown accuracy of these
results, however, here we also numerically investigate possible
systematic differences in the mass function with box size.

Over the redshifts and mass ranges probed in each of our simulation
boxes, we find no direct evidence for an error caused by finite volume
(at more than the $\sim 20\%$ level), as already emphasized previously
in \cite{Heitmann06}. (Overlapping box-size results over different
mass ranges are shown in Fig.~3 of~\citealt{Heitmann06}.)
Figure~\ref{plotten} shows the corresponding results in the present
work. This is not to say that there are no finite-volume effects (the
very high-mass tail in a given box must be biased low simply from
sampling considerations) but that their relative amplitude is
small. Below we discuss how to correct the mass function for finite
box size.

\subsubsection{Volume Corrections from Universality}
\label{sec_boxvolume_univ}

Let us first assume that mass function universality holds strictly, 
in other words, that for any initial condition the number of halos
can be described by a certain scaled mass function~(eq.~[\ref{fsigma}]) in
which $\sigma(M)$ is the variance of the top-hat-smoothed linear
density field. In the case of infinite simulation volume, $\sigma(M)$
is determined by equation~(\ref{sig}), and the mass function $\mf(M)$
of equation~(\ref{FM}) is
\begin{equation}
\mf(M) \equiv {dn \over d\log M}=\frac{\rho_b}{M}f(\sigma)
\frac{d\ln\sigma^{-1}}{d\log M}\ .
\label{eq_F_infinite}
\end{equation}
In an ensemble of finite-volume boxes, however, 
one necessarily measures a different quantity:  
\begin{equation}
\mf'(M') \equiv {dn' \over d\log M'}=\frac{\rho_b}{M'}f(\sigma')
\frac{d\ln\sigma'^{-1}}{d\log M'}\ .
\label{eq_F_finite}
\end{equation} 
Here $\sigma'(M')$ is determined by the (discrete) power spectrum of
the simulation ensemble, although if universality holds as assumed,
$f$ in equations~(\ref{eq_F_infinite}) and (\ref{eq_F_finite})
is the same function.

Since we are, in general, interested in the mass function
which corresponds to an infinite volume, we can then correct the data
obtained from our simulations as follows: 
for each box size we can define a function $M'(M)$ such that 
\begin{equation}
\sigma(M) \equiv \sigma'(M'(M)).
\label{sprime}
\end{equation}
Using equations (\ref{eq_F_infinite}) -- (\ref{sprime}), 
we determine $\mf(M)$ as    
\begin{equation}
\mf(M)=\mf'(M')\,\frac{dM'(M)}{dM}.
\end{equation}
Thus, the corrected number of halos in each bin is calculated as
\begin{equation}
dn = dn' \frac{M'}{M} \ .
\label{volcorr}
\end{equation}

The universality must eventually break down for sufficiently small
boxes or high accuracy because the nonlinear coupling of modes is more
complicated than that described by the smoothed variance.  This
violation can be partly corrected for by modifying the functional form of
$\sigma'(M')$. Therefore, we also explore other choices of
$\sigma'(M')$ which may better represent the mass function in the box.
To address this question we provide a short summary of the
Press-Schechter approach.

\subsubsection{Motivation from Isotropic Collapse}

We first consider the idealized case of a random isotropic
perturbation of pressureless matter and assume that the primordial
overdensity at the center of this perturbation has a Gaussian
probability distribution. The probability of local matter collapse at
the center is then fully determined by the local variance of the
primordial overdensity $\sigma^2$.  Consequently, for the isotropic
case the contribution of Fourier modes of various scales 
to the collapse probability is fully quantified by their
contribution to~$\sigma^2$.

To see this, consider the evolution of matter density $\rho_{\rm loc}$
at the center of the spherically symmetric density perturbation.  For
transparency of argument, let us focus on the evolution during the
matter-dominated era; it is straightforward to generalize the argument to
include a dark energy component $\rho_{\rm de}(z)$, homogeneous on the
length scales of interest, by a substitution $\rho_{\rm loc} \to
\rho_{\rm m,\,\rm loc}+\rho_{\rm de}$ in equations~(\ref{Freed_eq}) and
(\ref{q_collapse_def}).  By Birkhoff's law, the evolution of
$\rho_{\rm loc}$ and the central Hubble flow $H_{\rm loc}\equiv
\frac{1}{3}\bm{\nabla}\cdot\bm{v}_{\rm loc}$ are governed by the
closed set of the Friedmann and conservation equations,
\begin{eqnarray}
H_{\rm loc}^2&=&\frac{8\pi G \rho_{\rm loc}}{3}
              -\frac{\kappa}{a_{\rm loc}^2},
\label{Freed_eq}\\ 
a_{\rm loc}&\equiv&\left(\frac{\rho_0}{\rho_{\rm loc}}\right)^{1/3},~~ 
\frac{da_{\rm loc}}{dt}=H_{\rm loc}a_{\rm loc},
\label{a_loc}
\end{eqnarray}
where $\kappa$ is a constant
determined by the initial conditions, $\rho_0$ is arbitrary (e.g.,
$\rho_0=\left.\rho_b\right|_{z=0}$), and $t$ is the proper time.

The degree of nonlinear collapse at the center can be quantified by a
dimensionless parameter 
\begin{equation}
q\equiv 1 -\frac{3H_{\rm loc}^2}{8\pi G\rho_{\rm loc}}.
\label{q_collapse_def}
\end{equation}
First consider early times, when the evolution is linear, and let
$\rho_{\rm loc}=\rho_b(1+\delta)$. Then for the growing perturbation
modes during matter domination $H_{\rm loc}=\bar H(1-\delta/3)$.
Given these initial conditions, which set the initial $\rho_{\rm loc}$
and the constant $\kappa$ in equation~(\ref{Freed_eq}), the subsequent
evolutions of $\rho_{\rm loc}$, $H_{\rm loc}$, and therefore $q$ are
determined unambiguously.

During the linear evolution in the matter era $q=5\delta/3$ is small
and grows proportionally to the cosmological scale factor~$a$.  For
positive overdensity, nonlinear collapse begins when $q$ becomes of
order unity, reaching its maximal value $q=1$ when $H_{\rm loc}=0$,
and decreasing rapidly afterwards.  (We can observe the latter 
by rewriting eq.~[\ref{q_collapse_def}] as
\begin{equation}
q=\frac{3\kappa}{8\pi G a_{\rm loc}^2\rho_{\rm loc}} 
  \propto\rho_{\rm loc}^{-1/3},
\end{equation}
having applied eqs.~[\ref{Freed_eq}] and [\ref{a_loc}].) Nonlinear
collapse of matter at the center of the considered region can be said
to occur either when $q\rightarrow 0$ or when $q$ reaches a critical
``virialization'' value~$q_c$.

Now it is easy to argue that in the isotropic case the Press-Schechter
approach gives the true probability of the collapse, $P(q>q_c,z)$, for
a redshift $z$.  Indeed, the evolution of $q$ is set deterministically
by the primordial density perturbation at the center; for adiabatic
initial conditions specifically, it is set by the curvature
perturbation $\zeta$ at the center.  Since higher values of $\zeta$
lead to earlier collapse,
\begin{equation}
P(q>q_c,z)=P(\zeta>\zeta_c(z))=
\frac{1}{2}\mathop{\rm erfc}\left[\frac{\zeta_c(z)}{\sqrt2\,\sigma}\right],
\label{collapse_prob_isotrop}
\end{equation}
where the last equality uses the explicit form of $P(\zeta)$ as a
Gaussian distribution with a variance~$\sigma^2$.

\begin{figure}[t]
\begin{center}
\leavevmode\includegraphics[width=80mm]{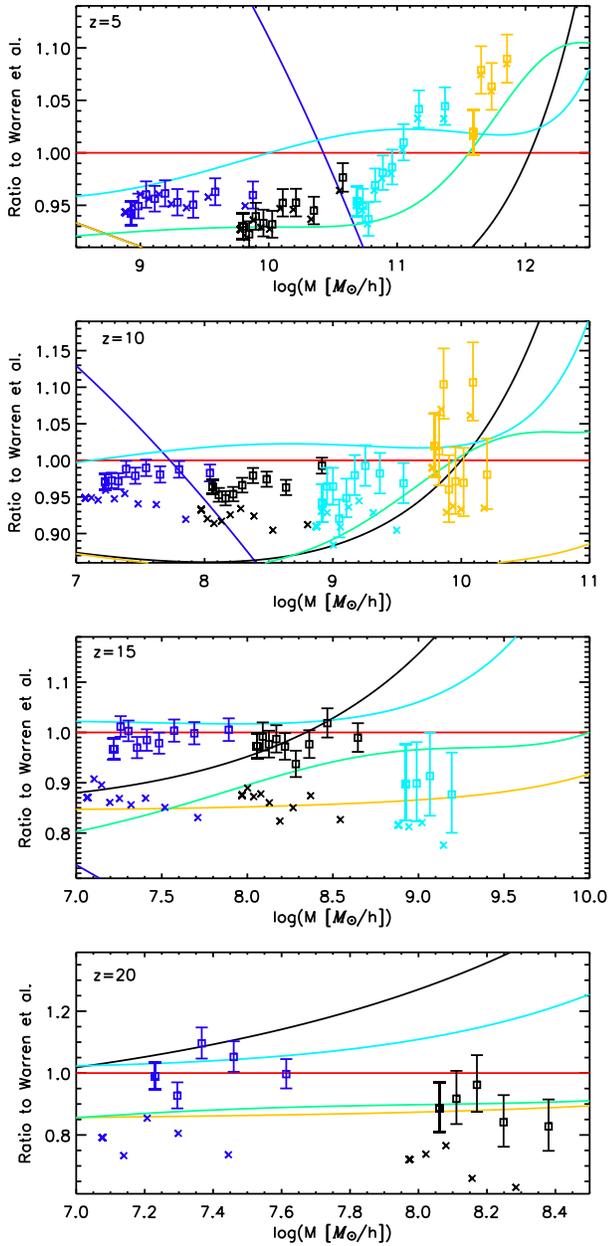}
\end{center}
\caption{Mass function data corrected for finite box volume
  by the extended Press-Schechter prescription
  of \S\ref{sec_boxvolume_num} (squares).
  We show the results as a ratio with respect to the
  Warren fit and follow the conventions of Fig.~\ref{plotten}.
  We also display the volume-uncorrected data (crosses).
  Note that the volume-corrected data join smoothly
  across the box-size boundaries.
  This box correction brings the results very close to
  universal behavior at high redshifts (see Fig.\ref{plotfifteen}).}
\label{plottwelve}
\end{figure}

If the considered isotropic distribution is confined by a (spherical)
boundary and $\sigma$ at the center is reduced by removal of
large-scale power, then equation~(\ref{collapse_prob_isotrop}) should
accurately describe the corresponding change of the collapse
probability. In numerical simulations, due to the imposition of
periodic boundary conditions, there is no power on scales larger than
the box size.  In this case the variance $\sigma$ should be specified
by the analogue of equation~(\ref{sig}) with the integral replaced by
a sum over discrete modes.

For the mass function~(eq.~[\ref{eq_F_infinite}]), a constant reduction of the
variance $\sigma^2(M)$ due to the removal of large-scale power leads
to a suppression of the mass function at the high-mass end and,
counterintuitively, a boost at the low-mass end. The latter is easily
understood as follows: The $\sigma$-dependent terms of
equation~(\ref{eq_F_infinite}),
\begin{equation}
f(\sigma)\frac{d\ln\sigma^{-1}}{d\log M}=\frac{d\rho(M)/\rho_b}{d\log M}, 
\end{equation}
give the fraction of the total matter density that belongs to the
halos of mass~$M$. When the variance is decreased by the box
boundaries, this fraction is boosted at low masses 
due to a shift of halo formation to an earlier stage, where a
larger fraction of matter is bound into low-mass objects.

\subsubsection{Numerical Results and Comparisons}
\label{sec_boxvolume_num}

Following the above intuition, we employ the extended Press-Schechter
formalism (\citealt{Bond91}) to correct for the missing fluctuation
variance on box scales. This formalism, while clearly inadequate at
various levels in describing halo formation in realistic simulations
(\citealt{Bond91}; \citealt{Katz93}; \citealt{White96}), has
nevertheless been very successful as a central engine in describing
the statistics of cosmological structure formation. As shown by
\cite{Mo96} using $N$-body simulations, the biasing of halos in a
spherical region with respect to the average mass overdensity in that
region is very well described by the extended Press-Schechter
approach. \cite{Barkana04} discussed the suppression of the halo mass
function in terms of this bias, and suggested a prescription for
adjusting large-volume mass function fits such as Warren or ST to
small boxes. Here we do not follow this path but directly work with
the numerical data by correcting the number of halos in each bin as in
equation~(\ref{volcorr}).

In the extended Press-Schechter scenario of halo formation, $\sigma'$
on the right-hand side of equation~(\ref{eq_F_finite}) would be
approximately connected with $\sigma$ via $\sigma'^2 = \sigma^2 -
\sigma_{R({\rm box})}^2$ \cite[]{Bond91}, where $\sigma_{R({\rm
    box})}^2$ is the variance of fluctuations in spheres that contain
the simulation volume. Since extended Press-Schechter theory is
derived for spherical regions, while our simulation boxes are cubes, we
define $R({\rm box})$ as the radius of a sphere enclosing the same
volume as in the simulations.

The action of this correction is shown in
Figure~\ref{plottwelve}. Finite-volume corrections are subdominant to
statistical error at $z=0$ and 5. At higher redshifts, the
corrections produce results that are consistent across box sizes,
i.e., that have no systematic shape changes or ``jumps'' across box
boundaries. Moreover, the action of the corrections is to bring the
simulation results closer to a universal behavior. We discuss
this aspect further below.

\begin{figure}[t]
\begin{center}
\leavevmode\includegraphics[width=80mm]{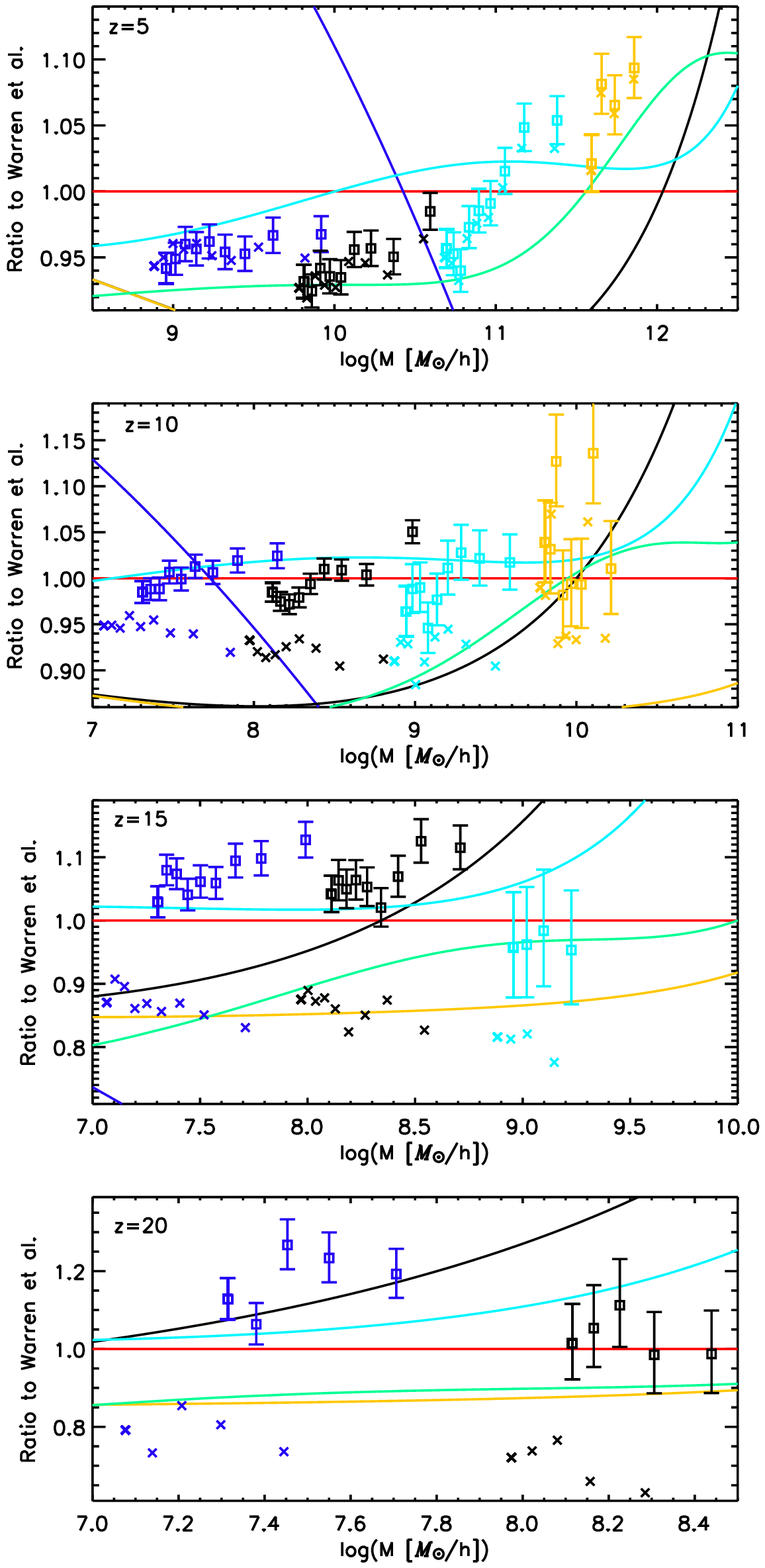}
\end{center}
\caption{}Mass function corrected for a finite box
  using the assumption of strict universality,
  as described in \S\ref{sec_boxvolume_univ} (squares).
  Again, we show uncorrected data as well (crosses), 
  and follow the conventions of Fig.~\ref{plotten}.
  This correction produces a clear systematic shift
  in the results across box boundaries.
\label{plotthirteen}
\end{figure}

For completeness, we mention two other approaches aimed at
box-adjusting the mass function.  The first (\citealt{Yoshida03c};
\citealt{Bagla06}) simply replaces the original mass variance 
(eq.~(\ref{sig})) with
\begin{equation}
\sigma^2_{\rm box}(M,z) = 
\frac{d^2(z)}{2\pi^2}\int^{\infty}_{2\pi/L}k^2P(k)W^2(k,M)dk\ ,
\label{sigbox}
\end{equation}
the lower cut-off arising from imposing 
periodic boundary conditions ($L$ is the box-size). (For enhanced
fidelity with simulations, the integral in eq.~[\ref{sigbox}]
goes to a sum over the simulation box modes.) This approach
basically assumes that $\sigma$ defined via an infrared cutoff is the
appropriate replacement for the infinite-volume mass variance. 
Figure~\ref{plotthirteen} shows the effect of this suggested
correction: At $z=0$ and 5 it is not noticeable, but at higher
redshifts the correction is significant relative to the accuracy
with which the binned mass function is determined. Furthermore, 
it exhibits 
systematic shape changes and offsets across boxes, in contrast to the
results shown in Figure~\ref{plottwelve}. For example, at $z=10$ the
corrected data at the crossover point between the 4 and 8$\,h^{-1}$Mpc
boxes ($\sim 10^8\,h^{-1}M_{\odot}$) have an offset of $5\%$. We
conclude that this approach is disfavored by our simulation results.

An alternative strategy is to estimate the mass variance from
each realization of $P(k)$ in the individual simulation boxes and to
treat every box individually, as done in \cite{Reed07}. This has in
fact two purposes: to compensate for the realization-to-realization
variation in density fluctuations 
(which could be a problem for small boxes) and
also to compensate for an overall suppression in the mass function as
discussed above. The disadvantage is that each 
of many realizations now has a different
$\sigma(M)$ for a given value of $M$.

\subsection{Mass Function Universality}

Finally, we investigate the universality of the mass function found by
Jenkins.  Approximate universality is expected from the
analytic arguments of PS and the extended,
excursion-set formulation of Bond et al. (1991).  The universal
behavior of halo formation persists even in the model of ellipsoidal
collapse of ST, in which the predicted mass
function is no longer of the PS form.  On the other hand, the
universality cannot be exact if the nonlinear interactions of different
scales are fully accounted for: The nonlinear evolution that leads to
the formation of halos of mass $M$ must involve multiple degrees of
freedom that are described by more parameters than the overall
variance of the primordial overdensity smoothed by a top-hat filter
$W(r,M)$.  The universality is expected to be violated at sufficiently
high resolution of the mass function even in the PS-type spherical
collapse model: It is more reasonable to represent the probability of
the collapse not by a fraction of particles at the center of spheres
enclosing a mass~$M$ but by any fraction of particles belonging to such spheres
\citep{beta106}.  The improved mass-function derived from this argument
deviates somewhat from a universal form \citep{beta206}.

\begin{figure}[t]
\begin{center}
\leavevmode\includegraphics[width=7.5cm]{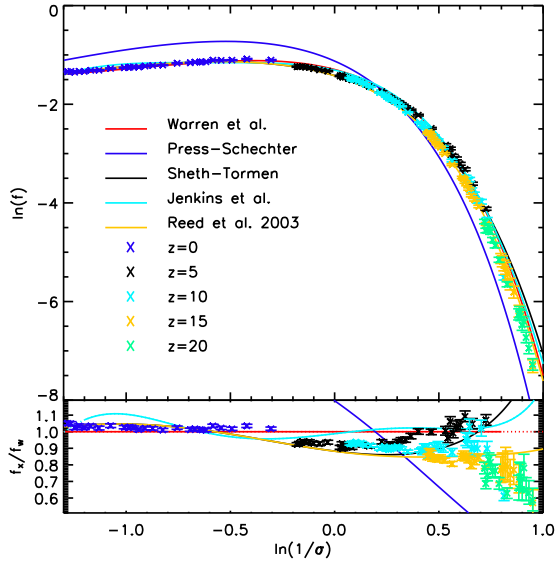}
\end{center}
\caption{Scaled differential mass function from all simulations,
  prior to applying finite-volume corrections. Fits
  shown are Warren (red), PS (dark blue), ST (black), Jenkins (light
  blue), and Reed et al.~(2003) (yellow). Dashed lines denote an
  extrapolation beyond the original fitting range. The bottom panel
  shows the ratio relative to the Warren fit. The failure of the
  different redshift results to lie on top of each other at small
  values of $\sigma$ indicate a possible violation of universality.}
\label{plotfourteen}
\end{figure}

\begin{figure}[t]

\begin{center}
\leavevmode\includegraphics[width=7.5cm]{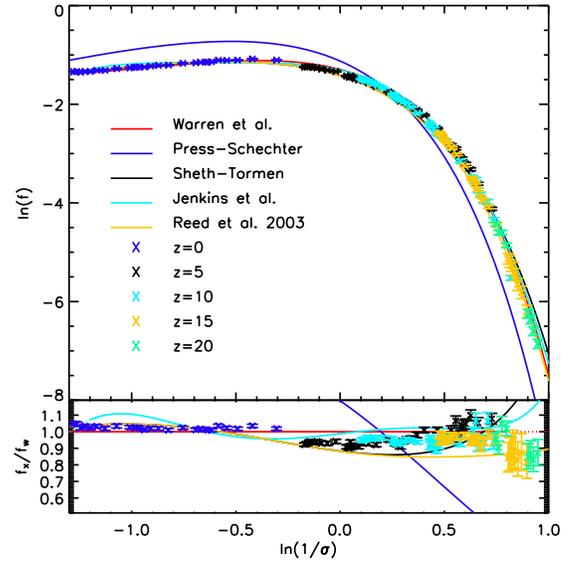}
\end{center}
\caption{Volume-corrected scaled differential mass function following
  Fig.\ref{plotfourteen}. Note the significantly improved agreement
  with universal behavior (overlapping results beyond
  $\ln\sigma^{-1}\sim0.3$).}
\label{plotfifteen}
\end{figure}

To investigate the extent our numerical simulations are consistent
with universality, we combine our results for $f(\sigma,z)$ as a
function of the variance $\sigma^{-1}$ from the entire simulation set
in one single curve at various redshifts.  This curve is expected to
be independent of redshift if universality holds.  We display the
results in Figure~\ref{plotfourteen} for the raw data and in
Figure~\ref{plotfifteen} for the same data after applying the volume
corrections discussed earlier.

In the raw data of Figure~\ref{plotfourteen}, the agreement with the
various fits is quite tight (except for PS) until
$\ln\sigma^{-1}>0.3$. Beyond this point, the multiple-redshift
simulation results do not lie on top of each other; in the absence of
any possible systematic deviation, this would denote a failure of the
universality of the FOF, $b=0.2$ mass function at small
$\sigma$. Note also that beyond this point the ST and Jenkins fits
have a steeply rising asymptotic behavior (relative to the Warren
fit). The Reed et al.~(2003) fit, meant to be valid over the range $-1.7\le
\ln \sigma^{-1} \le 0.9$, is in better agreement with our results, to
the extent that a single fit can be overlaid on the data.

The ostensible violation of universality seen above is small, however,
and subject to a systematic correction due to the finite simulation
volume(s). On applying the correction discussed in Section~5.6.3, we
obtain the results shown in Figure~\ref{plotfifteen}, the key
difference being that beyond $\ln\sigma^{-1}>0.3$ the multiple-redshift 
simulation results now lie on top of each other and, within
the statistical resolution of our simulations, are consistent with
universal behavior. Specifically, we do not observe the sort of
violation reported by \cite{Reed07} at high redshifts. This could be
due to several factors. The finite-sampling FOF mass correction and
the finite-volume corrections we employ are different from those of
\cite{Reed07} and the boxes we use at high redshifts are
significantly larger. We note also that the difference between the
Warren fit and the $z$-dependent fit of \cite{Reed07} does not appear
to be statistically very significant given their data. 

\section{Conclusions and Discussion}
\label{conclusion}

We have investigated the halo mass function from $N$-body simulations
over a large mass and redshift range.  A suite of 60
overlapping-volume simulations with box sizes ranging from
4 to 256$\,h^{-1}$Mpc allowed us to cover the halo mass
range from 10$^7$ to 10$^{13.5}\,h^{-1}M_\odot$ and an effective
redshift range from $z=0$ to 20.

In order to reconcile conflicting results for the mass function at
high redshifts, as well as to investigate the reality of the breakdown
of the universality of the mass function, we have studied various
sources of error in $N$-body computations of the mass function. A set
of error control criteria need to be satisfied in order to obtain
accurate mass functions. These simple criteria include an estimate for
the necessary starting redshift, for the required mass and force
resolution to resolve the halos of interest at a certain mass and
redshift, and for the number of time steps.

The criteria for the initial redshift appear to be particularly
restrictive.  For small boxes, commonly used in the study of the
formation of the first objects in the Universe, significantly higher
initial redshifts are required than is the normal practice.  A
violation of this criterion leads to a strong suppression of the mass
function, most severe at high redshifts. Recent results by other
groups may be contaminated due to a violation of this requirement; a
careful re-analysis of small-box simulations is apparently indicated.

The force resolution criterion is especially useful for grid codes, PM
as well as adaptive mesh. The mass function can be obtained reliably
from PM codes down to small-mass and up to high-mass halos provided the
halos are adequately resolved.  The resolution criterion is also very
useful in setting refinement levels for adaptive mesh refinement (AMR)
codes.  As shown recently (O'Shea et al.\ 2005; Heitmann et al.\ 2005, 
2007) 
the mass function from AMR codes is suppressed at the low-mass end if
the base refinement level is too coarse. A more detailed analysis on
how to incorporate our results to improve the efficiency of AMR codes
is underway.

The results for the required number of time steps to resolve the mass
functions is somewhat surprising. The halo mass function appears to be
very robust with respect to the number of time steps chosen to follow
the evolution, even though the inner structure of the halos will
certainly not be correct. Even a small number of time steps is
sufficient to obtain a close-to-correct mass function at $z=0$. This
considerably simplifies the study of the mass function and its
evolution.

Since finite-volume effects can also lead to a suppression of the mass
function, we have tried to minimize the importance of these effects by
avoiding too-small box sizes, by using overlapping boxes, and by
restricting the mass range investigated in a given
box size. In addition, we have found that a box-size correction
motivated by the extended Press-Schechter formalism for the mass
variance appears to give consistent results when applied to our
multiple-box simulation ensembles.

We now briefly comment on results found previously by other
groups. Jang-Condell \& Hernquist (2001) find good agreement with the
PS fit at $z=10$ for a mass range $4 \times 10^{5} - 4 \times 10^{8}
\,h^{-1}M_\odot$. The crossover of PS with the more accurate fits at
$z=10$ takes place in exactly this region (see Figs.~\ref{plotone}
and \ref{plotten}).  Therefore, all fits are very close, and the
mass function from a single 1 Mpc box at a single redshift as shown in
Jang-Condell \& Hernquist (2001) cannot distinguish between them.

As mentioned earlier in \S\ref{evolreview}, good agreement with
the PS result has been reported at high redshifts (some results being
even lower than PS) by several other groups (Yoshida et al.  2003a,
2003b, 2003c; \citealt{Cen04};
\citealt{Trac06}). The simulations of \cite{Cen04} and \cite{Trac06}
were started at $z_{\rm in}\sim 50$, substantially below the starting
redshift that would be suggested by our work. The very large number of 
particles in the \cite{Trac06} simulation 
requires a high starting redshift 
(Fig.~\ref{plotfive}). Therefore, the
depressed mass function results of these simulations are very
consistent with a too-low initial redshift. (\cite{Trac06} have
recently rerun their simulations with a much higher initial redshift
[$z=300$], and now find results consistent with ours.) The initial
particle density of the \cite{Iliev05} simulations is very close to
that of our 16$\,h^{-1}$Mpc box, in which case also a high redshift
start is indicated (we used $z_{\rm in}=200$). Finally, the initial
redshift of the Yoshida et al. papers, $z_{\rm in}=100$, for boxes of
size $\sim 1\,h^{-1}$Mpc, also appears to be significantly on the low
side.

We have compared our simulation results for the mass function with
various fitting functions commonly used in the literature. The
recently introduced ($z=0$) fit of Warren leads to good
agreement (at the 20\% level with no volume correction, and at the 5\% level
with volume correction) at all masses and all redshifts we
considered. Other modern fits, such as Reed et al.~(2003, 2007), also
lie within this range. These fits do not suffer from the
overprediction of large halos at high redshifts observed for the ST
fit.  The PS fit performs poorly over almost all the considered mass
and redshift ranges, at certain points falling below the simulations
by as much as an order of magnitude.

The evolution of the mass function can be used to test the
(approximate) universality of the FOF, $b=0.2$ mass function. At low
redshifts our data are in good agreement with those of \cite{Reed07} (at
$z=5$), finding a (possible) mild redshift dependence (at the 10\%
level). At higher redshifts, however, we find that volume corrections
are important to the extent that little statistically significant
evidence for breakdown of universality remains in our mass function
data. A full theoretical understanding of this very interesting result
remains to be elucidated.

We have made no attempt to provide a fitting function for our data due
to several reasons. First, the current simulation state of the art has
not reached the point that one can be confident of percent-level
agreement between results from different simulations even in regimes
that are not statistics-dominated \cite[]{Heitmann07}. Second, 
simulations have not
sufficiently explored the extent to which universal forms for the mass
function are indeed applicable as cosmological parameters are
systematically varied. Third, absent even a compelling
phenomenological motivation for the choice of fitting functions, there
is an inherent arbitrariness in the entire procedure. Finally, it is
not clear how to connect the FOF mass function to observations. In
general, tying together mass-observable relations requires close
coupling of simulations and observational strategies. In studies of
cosmological parameter estimation, we support working directly with
simulations rather than with derived quantities, which would add another layer
of possible systematic error. Because observations already
significantly constrain the parametric range, and are a smooth
function of the parameters, this approach is quite viable in practice
(\citealt{coscal1}; \citealt{coscal2}).

\acknowledgements

We thank Kevork Abazajian, Nick Gnedin, Daniel Holz, Lam Hui, Gerard
Jungman, Savvas Koushiappas, Andrey Kravtsov, Steve Myers, Ken
Nagamine, Brian O'Shea, Sergei Shandarin, Ravi Sheth, Mike Warren, and
Simon White for helpful discussions. We are particularly grateful to
Adam Lidz and Darren Reed for their help, many discussions, and 
sharing their results with us. We are indebted to Mike Warren for use
of his parallel FOF halo finder.  The authors acknowledge support from
the Institute of Geophysics and Planetary Physics at Los Alamos
National Laboratory. S. B., S. H., and K. H. acknowledge support from the
Department of Energy via the Laboratory-Directed Research and Development 
program of Los Alamos.  P. M. R.
acknowledges support from the University of Illinois at
Urbana-Champaign and the National Center for Supercomputing
Applications.  S. H., K. H., and P. M. R.  acknowledge the hospitality of
the Aspen Center for Physics, where part of this work was carried out.
The calculations described herein were performed using the
computational resources of Los Alamos National Laboratory.  A special
acknowledgement is due to supercomputing time awarded to us under the
LANL Institutional Computing Initiative. P. M. R. and Z. L. also acknowledge
support under a Presidential Early Career Award from the
US Department of Energy, Lawrence Livermore National Laboratory
(contract B532720).

\appendix
\label{append}

In this Appendix we discuss in detail previous results on the mass
function at high redshift. As explained in the main paper, these
results are often contradictory. We structure our discussion with
respect to the physical volume simulated.

\subsection{Small-Volume Simulations} Small-box simulations of
side $\sim 1\,h^{-1}$Mpc have been performed by several groups. Using a
treecode with softening length $0.4\,h^{-1}$kpc, and a 1$\,h^{-1}$Mpc
box with $128^3$ particles, \cite{Jangcondell01} evolved their
simulation from $z_{\rm in}=100$ to $z=10$. With a halo finder that
combined overdensity criteria with an FOF algorithm, the mass function
was determined over the range $10^{5.5} - 10^{8.1}\,h^{-1}M_\odot$,
keeping halos with as few as eight particles.  At $z=10$ they found
``remarkably close agreement'' with the PS fit but did not quantify
the agreement explicitly.

In a series of papers, Yoshida et al. ran simulations with similar box
sizes as above, most including the effects of gas dynamics.  The
simulations were performed with the TreePM/smoothed particle hydrodynamics code
GADGET-II~\cite[]{Springel05b} and followed the evolution of 2$\times$
324$^3$ particles (324$^3$ in the case of dark matter only), covering
a halo mass range of 10$^5$-10$^{7.5}M_\odot$.  All simulations were
started at $z_{\rm in}=100$ from ``glass'' initial conditions (Baugh 
et al.~1995; White 1996), in contrast to the
grid-based initial conditions used here. The focus of
\cite{Yoshida03a} was the origin of primordial star-forming clouds.
As part of that investigation, a dark-matter-only simulation in a
1.6$\,h^{-1}$Mpc box was carried out. The halo density results for
$z=20$ to 32 lay systematically below the PS prediction, with the
discrepancy being worse at high redshifts.  The authors argued that
this low abundance of halos was (possibly) due to finite-box-size
effects.  In \cite{Yoshida03b}, the mass function at $z=20$ for a warm
dark matter model was compared with CDM, with the simulation set up being
very similar to that of \cite{Yoshida03a}, a 1 Mpc box started at $z=100$.
The results obtained were also similar; at $z=20$ the CDM mass
function was in good agreement with the PS fit.  In a third paper,
\cite{Yoshida03c}, a running spectral index was considered. Here
results for a standard CDM mass function for a 1~Mpc box were given,
this time at $z=17$ and 22.  Consistent with their previous
results, they found good agreement with PS at these redshifts.  (The
FOF linking length used in the last paper was $b=0.2$, while in the
first two papers $b=0.164$ was chosen.  This did not appear to make
much of a difference, however.)  These papers do not quantitatively
compare the numerical mass function to the PS fit. (In contrast to
these findings, a recent 1~Mpc box GADGET-II simulation with $z_{\rm
  in}\sim 120$ has been performed by \cite{Maio06} who find good
agreement with the Warren fit as extrapolated by linear theory -- in
clear disagreement with PS.)

A similar strategy was followed in \cite{Cen04} who investigated dark
matter halos in a mass range of $10^{6.5}$ to
$10^9\,h^{-1} M_\odot$, using a TreePM code (Xu 1995; Bode et al.~2000). The
box size was taken to be $4\,h^{-1}$Mpc, the softening length was set at
$0.14\,h^{-1}$kpc, $512^3$ particles were used, and the simulations had
a starting redshift of $z_{\rm in}=53$. Halos were identified using
the overdensity scheme DENMAX \cite[]{Bertschinger91}.  Among other
quantities, they studied the mass function between $z=11$ and 6
and found that the PS function ``provides a good fit'' but without
explicit quantification.

Overall, these small-box simulations, run with different codes and
different halo finders, all found a ``depressed'' mass function
(see Fig.~\ref{plotone}), consistent with PS and deviating very
significantly from the predictions of the more modern fitting
forms. In contrast, other simulations also using small boxes have
come to quite different conclusions. For example, in \cite{Reed07}, a
large suite of different box sizes and simulations was used to cover
the mass range between 10$^5$ and 10$^{11.5}
\,h^{-1}M_\odot$ at high redshift. The smallest boxes considered in
this study were 1~$\,h^{-1}$Mpc on a side. The authors studied the halo
mass function at redshifts out to $z=30$, implementing a correction
scheme to account for finite-box effects, as discussed in more detail
below. Overall, \cite{Reed07} confirmed previous results as found by
\cite{Reed03} and \cite{Heitmann06}: PS underestimates the mass
function considerably (by at least a factor of 5 at high redshift
and high masses), and ST overpredicts the halo abundance at high
redshift.

\subsection{ Large-Volume Simulations}
The large-box strategy is exemplified by a recent dark matter
simulation with the GADGET-II code \cite[]{Springel05a}. The evolution
of 2160$^3$ particles in a $500 \,h^{-1}$Mpc box was followed from
$z_{\rm in}=127$ until $z=0$. The softening length was $5\,h^{-1}$kpc.
The high mass and force resolution was sufficient to study the mass
function reliably down to a redshift of $z=10$, covering a mass range
of $10^{10}$ to $10^{16} \,h^{-1}M_\odot$, with halos being
identified by a standard FOF algorithm with $b=0.2$.  The results
are consistent with the Jenkins fit, even though the mass function
points at redshifts $z=1.5$, 3.06, and 5.72 are slightly higher
than the Jenkins fit and slightly lower for $z=10$. No residuals were
shown nor quantitative statements made.

In two recent papers, \cite{Iliev05} and \cite{Zahn06} investigated
cosmic reionization, providing mass function results at high redshift
as part of this work. \cite{Iliev05} ran a PM simulation with
PMFAST~(Merz et al.~2005) in a 100$\,h^{-1}$Mpc box with 1624$^3$
particles on a 3248$^3$ mesh. They present results for the mass
function at redshifts between $z=6$ and 18.5, using a spherical
overdensity halo finder. At lower redshifts they find good agreement
with ST, and at high redshift ($z>10$) the results are closer to PS
(because of their limited mass range, a more quantitative statement is
difficult to make). \cite{Zahn06} ran a 1024$^3$ particle simulation
(dark matter only) in a 65.6$\,h^{-1}$Mpc box with GADGET-II and
analyzed the FOF, $b=0.2$ mass function out to $z=20$. Between $z=6$
and 14 they found good agreement with ST in the mass range of 10$^9$
to $10^{12}M_\odot$. At $z=20$ they found that the simulation
results were below ST but above PS, in relatively good agreement with
the recent findings of~\cite{Heitmann06} and \cite{Reed07}.

\subsection{ Medium Volume Simulations}

\cite{Reed03} chose a compromise between the large- and small-box
strategies by picking a 50$\,h^{-1}$Mpc box sampled with 432$^3$
particles. The tree code PKDGRAV was used to evolve the simulation
from different starting redshifts between $z_{\rm in}=139$ and 
69 until $z=0$.  The smallest halo contained 75 particles,
leading to a mass range of roughly 10$^{10}$ to
$10^{14.5}\,h^{-1}M_\odot$.  Good agreement (better than 10\%) was
found with the ST fit up to $z\simeq 10$.  For higher redshifts, the
ST fit overpredicted the number of halos, up to 50\% at $z=15$.  At
this high redshift, statistics were lacking, and the resolution was not
sufficient to resolve very small halos. A more recent 50$\,h^{-1}$Mpc
simulation with PMFAST with $z_{\rm in}=60$ has been carried out by
\cite{Trac06} using a spherical overdensity definition of halo
mass. In this work, the mass function, in the redshift range $6<z<15$,
is found to be in very good agreement with PS, in gross contradiction
with the results of most of the other simulations mentioned above. (This
contradiction has recently been resolved by rerunning their simulation
with $z_{in}=300$ and identifying halos with a $b=0.2$ FOF finder.)

\end{document}